\documentclass[journal, twocolumn, 11pt]{IEEEtran}
\usepackage{graphicx}
\usepackage{epstopdf}
\usepackage{amsfonts}
\usepackage{multirow}
\usepackage{color}
\usepackage{comment}
\usepackage{algorithm}
\usepackage{algorithmic}
\usepackage{setspace}
\usepackage{hyperref}
\usepackage{breakurl}
\usepackage{amssymb}
\usepackage{enumitem}
\usepackage{amsmath}
\begin{document}

\title{{\LARGE{A Security Credential Management System for V2X Communications}}}

\author{
\IEEEauthorblockN{
        Benedikt Brecht\IEEEauthorrefmark{5},
        Dean Therriault\IEEEauthorrefmark{6},
        Andr\'e Weimerskirch\IEEEauthorrefmark{2},
        William Whyte\IEEEauthorrefmark{1},
        Virendra Kumar\IEEEauthorrefmark{1},
        Thorsten Hehn\IEEEauthorrefmark{3},
        Roy Goudy\IEEEauthorrefmark{4}
}
\\

\IEEEauthorblockA{
        \IEEEauthorrefmark{5}Benedikt.Brecht@vw.com
}
\\

\IEEEauthorblockA{
        \IEEEauthorrefmark{6}dean.therriault@gm.com
}
\\

\IEEEauthorblockA{
        \IEEEauthorrefmark{2}aweimerskirch@lear.com
}
\\

\IEEEauthorblockA{
        \IEEEauthorrefmark{1}\{wwhyte, vkumar\}@onboardsecurity.com
}
\\

\IEEEauthorblockA{
        \IEEEauthorrefmark{3}thehn@gmx.de
}
\\

\IEEEauthorblockA{
        \IEEEauthorrefmark{4}goudyr1@nrd.nissan-usa.com
}
}

\maketitle

\graphicspath{{graphics/}}
\definecolor{grey}{rgb}{0.4, 0.4, 0.4}

\newcommand{\CertID}{\mathrm{CertID}}
\renewcommand{\H}{\mathrm{H}}
\newcommand{\E}{\mathrm{E}}
\newcommand{\EXOR}{\mathrm{XOR}}
\newcommand{\ls}{\mathrm{ls}}
\newcommand{\gs}{\mathrm{gls}}
\newcommand{\laid}{\mathrm{la\_id}}
\newcommand{\plv}{\mathrm{plv}}
\newcommand{\gplv}{\mathrm{gplv}}
\newcommand{\lci}{\mathrm{lci}}
\newcommand{\lv}{\mathrm{lv}}
\newcommand{\glv}{\mathrm{glv}}

%
%

\begin{abstract}
The US Department of Transportation (USDOT) issued a proposed rule on January 12th,
2017 to mandate vehicle-to-vehicle (V2V) safety communications in light vehicles in
the US. Cybersecurity and privacy are major challenges for such a deployment. The authors
present a Security Credential Management System (SCMS) for vehicle-to-everything (V2X)
communications in this paper, which has been developed by the Crash Avoidance Metrics Partners LLC
(CAMP) under a Cooperative Agreement with the USDOT. This system design is currently
transitioning from research to Proof-of-Concept, and is a leading candidate to
support the establishment of a nationwide Public Key Infrastructure (PKI) for V2X 
security. It issues digital certificates to participating vehicles
and infrastructure nodes for trustworthy communications among them,
which is necessary for safety and mobility applications that are based on V2X
communications. The SCMS supports four main use cases, namely
bootstrapping, certificate provisioning, misbehavior reporting and
revocation. The main design goal is to provide both security and
privacy to the largest extent reasonable and possible. To achieve
a reasonable level of privacy in this context, vehicles are issued pseudonym certificates,
and the generation and provisioning of those certificates are divided
among multiple organizations. Given the large number of pseudonym
certificates per vehicle, one of the main challenges is to facilitate efficient
revocation of misbehaving or malfunctioning vehicles, while preserving
privacy against attacks from insiders. The proposed SCMS supports all
identified V2X use-cases and certificate types necessary for V2X communication security.

This paper is based upon work supported by the U.S. Department of Transportation. Any opinions, findings, and conclusions or recommendations expressed in this publication are those of the Authors ("we") and do not necessarily reflect the view of the U.S. Department of Transportation.
\end{abstract}

%
%

\section{Introduction}
\label{sec:introduction}

Vehicle-to-Vehicle (V2V) communications between nearby vehicles in the
form of continuous broadcast of Basic Safety Messages (BSMs) has the
potential to reduce unimpaired vehicle crashes by 80\% through active
safety applications~\cite{usdot-cv}.
Following a series of field
operational tests, the US Department of Transportation (USDOT)
issued a proposed rule  on January 12th, 2017 to mandate the inclusion of V2V
technology in light vehicles in the US~\cite{usdot-2017}.
Vehicles will broadcast BSMs up to ten times per second to support V2V
safety applications. BSMs include the senders' time, position,
speed, path history and other relevant information, and are digitally
signed.  The receiver evaluates each message, verifies the signature,
and then decides whether a warning needs to be displayed
to the driver. The correctness and reliability of BSMs are of prime
importance as they directly affect the effectiveness of
safety applications based on them. To prevent an attacker from
inserting false messages, the sending
vehicles digitally sign each BSM, and the receiving vehicles verify
the signature before acting on
it. This approach has been recommended by many different studies of
the system in both Europe and America.~\cite{c2c11,etsi-tr,etsi-ts1,etsi-ts2,sevecom,vscc,1609.2}.

A Public-Key Infrastructure (PKI) that facilitates and manages digital
certificates is necessary for building trust among participants and
for proper functioning of the system. The Security Credential
Management System (SCMS) proposed in this paper implements a PKI with unique features. This
SCMS design is currently a leading candidate for the V2X security
backend in the US. It is distinguished from a traditional PKI in several
aspects. The two most important aspects being its size (i.e., the number
of devices that it supports) and the balance among security, privacy,
and efficiency. At its full capacity, it will issue approximately
$300$ billion certificates per year\footnote{This number may be even
greater if pedestrian and cyclist-borne units become part of the
system.} for $300$ million vehicles. To the best of our knowledge,
the PKI whose root is run by the Europay-Mastercard-Visa Consortium
(EMVCo) is the largest extant PKI in the world and issues in the 
single-digit billions of certificates per year~\cite{emvco}, while the largest current
government-run PKI, deployed by the Defense Information Systems Agency for the
Common Access Cards program~\cite{wiki:pki}, is several orders of magnitude smaller and issues
under $10$ million certificates per year. 
At the core of its design, the proposed SCMS
has several novel cryptographic constructs to provide a high level of
security and privacy to the users while keeping the system very
efficient. As a result, the presented SCMS design is significantly
different from any previously implemented PKI. However, it is somewhat
similar to the design of the European C2X PKI~\cite{c2c11}. The SCMS
was designed with privacy (both against SCMS insiders and outsiders)
being the highest priority. The SCMS design also provides efficient
methods for requesting certificates and handling revocation. An early
version of the proposed SCMS has already been implemented, operated
and tested in the Safety Pilot Model Deployment~\cite{safety-pilot},
and the design presented here has been specified and
implemented in the SCMS Proof-of-Concept (PoC) Project and is going to be deployed for 
the USDOT's Connected Vehicle Pilot Program (CV Pilot Program)~\cite{usdot-2017}.
Besides V2V safety applications, there will also be
vehicle-to-infrastructure and infrastructure-to-vehicle (V2I / I2V)
applications\footnote{For the sake of a compact representation, we
will denote V2I / I2V applications as V2I applications} to support
safety, mobility and environmental applications. While V2V safety
applications are the current focus of research and deployment in the
US, it is essential for the SCMS to also be able to cover V2I
applications. All applications in the scope of the
current research and the discussion of this paper are limited to
single-hop communication. A wide variety of V2I applications were analyzed and categorized, and
the SCMS design was updated to support all identified V2I application
categories. This includes infrastructure-originating broadcast
messages (e.g.,\ traffic light announcements) as well as service
announcement and provisioning (e.g.,\ Internet access). The first
application category requires that road-side equipment (RSE)
authenticates broadcast messages, whereas the second application
category requires that on-board equipment (OBE) in the vehicle
establishes a communication channel to the RSE.

The interface specification of the current version of the SCMS is available
from~\cite{scms-interface-spec}. The goal of this paper is to provide
an overview of the features of the SCMS along with rigorous
arguments for the appropriateness of the design decisions made.

%
%

\section{SCMS Design Overview}
\label{sec:overview}
\begin{figure*}
\begin{center}
\scalebox{0.65}{
\input{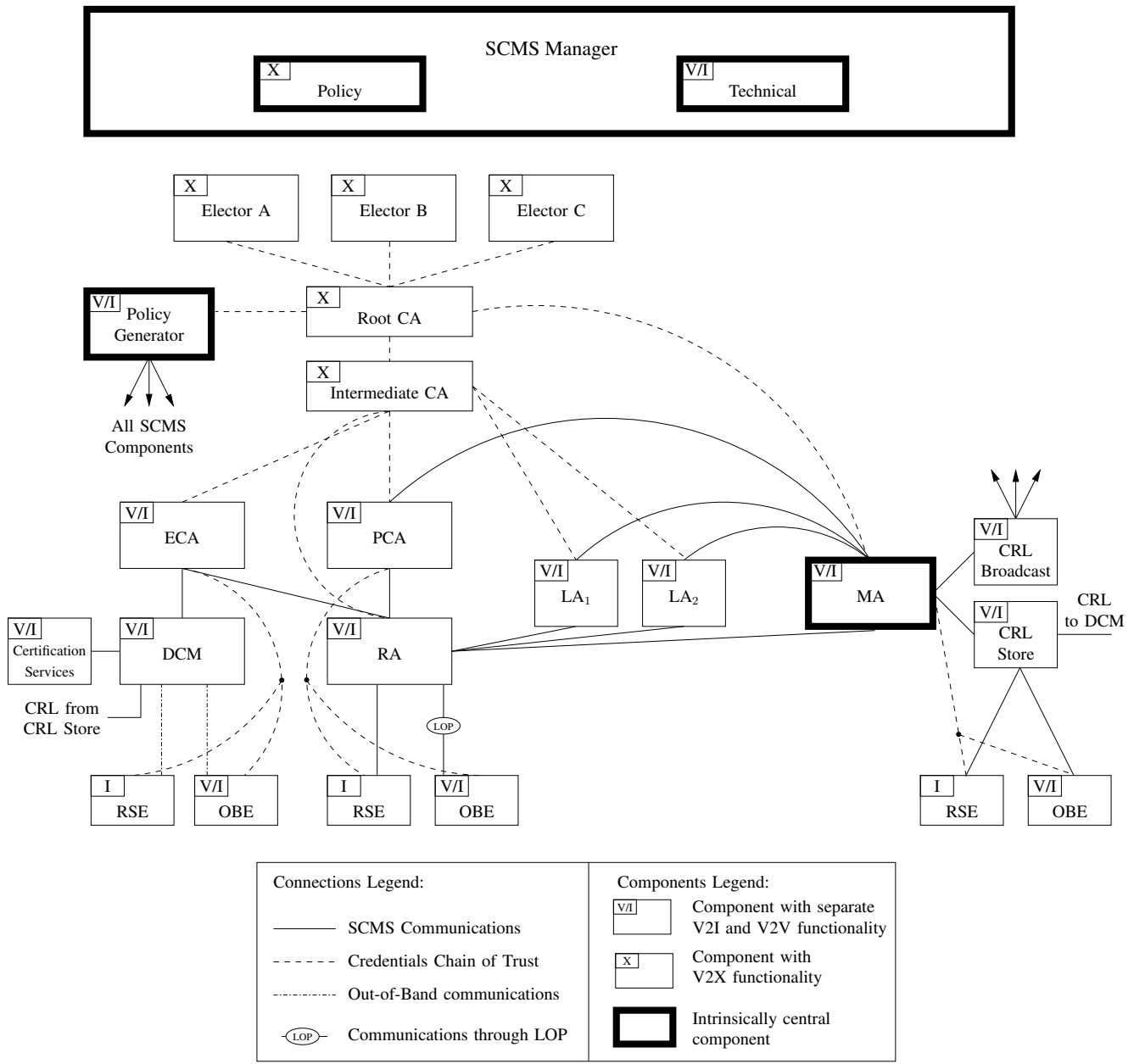}
}
\caption{\label{fig:overview}SCMS Architecture Overview
}
\end{center}
\end{figure*}

In this section, we present the design of the SCMS by briefly
explaining its components and then discussing the rationale behind them.
We say that an SCMS component is {\em intrinsically-central},
if it can have exactly one distinct instance for proper functioning. A
%
%
component is {\em central}, if it is chosen to have exactly one
distinct instance in the considered instantiation of the
system. Distinct instances of a component have different identifiers
and do not share cryptographic materials. While there is only one SCMS,
components that are not central can have multiple instances.
%
%
It is assumed that all components support
load balancing if needed. Figure~\ref{fig:overview} gives an overview
of the overall system architecture. The lines connecting different
SCMS components in Figure~\ref{fig:overview} are relationship lines, meaning that in
at least one of the use cases, one component sends information or
certificates to the other. The SCMS was originally designed for
V2V use cases~\cite{wwkh13}, but later extended to support V2I as well.
We will later present the different types of
certificates required to support both V2V and V2I applications.
In the SCMS, there are components
that are included merely for V2V services (e.g., Linkage Authority),
for generic V2X services (e.g., Intermediate CA), and for combined V2V
and V2I services with separate features for V2V and V2I, respectively
(e.g., Pseudonym CA). Additionally, there is a dotted line connecting a
device to the SCMS to indicate out-of-band secure communication. 
The SCMS can be simplified at the expense of losing some flexibility,
e.g.,\ by making all the components {\em central}.

\subsection{Threat Models and Application Concepts}
An (unpublished) risk assessment was performed for V2V safety applications, which at
the moment are being considered for driver notification only (i.e.,\ not
for control, and not for traffic management or other mobility applications). 
For instance, a V2V safety application might be designed
to provide a warning to the driver in case of an imminent forward
collision but such an application will not be able to control the car
based on the wireless Dedicated Short Range Communication (DSRC)
message input alone. It is assumed that the cryptographic private keys
and other security- and privacy-sensitive materials are stored and
used securely in secure hardware.  The risk assessment
concluded the major risk to be that of users deactivating the DSRC
system in their cars in case they receive too many false warnings due
to bogus messages broadcast by malicious devices. Since V2V is a
collaborative technology that requires participation from a majority
of vehicles, the system can be considered a failure if too many nodes
were to deactivate for any reason. 

The risk assessment also
concluded that there is no safety-of-life risk from security hacking
attacks on the SCMS, since the result of hacking the SCMS is that an attacker
can either send false messages or create a denial of service attack on the
system (e.g., by widespread illegitimate revocation of devices). Neither of 
these hacks are a direct risk to safety-of-life in the sense of making collisions more 
likely than they would be in the absence of the system given the above assumption 
about driver notification applications. However, these hacks would 
clearly reduce the safety-of-life benefits of the system, and as such, the security of 
the SCMS components is very important to the system's success.
%
%

Cybersecurity aspects of the vehicle on-board DSRC computing platform and the SCMS components
are excluded in this paper. However a secure design and separation of safety-critical systems from
the rest of the vehicle systems are needed. We believe that the cybersecurity concerns of DSRC units
are similar to other wireless interfaces such as cellular and Bluetooth. 

Furthermore, the risk assessment concluded that there are risks to privacy
from security attacks by SCMS outsiders as well as insiders. It was thus 
concluded that the SCMS must counter / mitigate the following types of attacks:

\begin{itemize}
\item Attacks on end-users' privacy from SCMS insiders
\item Attacks on end-users' privacy from outside the SCMS
\item Authenticated bogus messages leading to false warnings
\end{itemize}

We address the first two items listed above by what we call ``Privacy by Design.''
The third item is addressed by misbehavior detection (to identify
misbehaving and malicious devices) and efficient revocation (to reject messages from revoked devices)
to avoid / minimize the harm to the system's trustworthiness. 
While the future risk assessments are expected to provide a
better understanding of the properties and parameters of the
system, the types of certificates to be issued by the SCMS are
considered to be stable at the moment. For the design of the
SCMS, we assume that today's cryptographic mechanisms are
acceptably secure. This assumption becomes incorrect if 
quantum computers of sufficient scale are developed as
they could break the Elliptic Curve Digital 
Signature Algorithm for all practical curve sizes. The design is
modular and flexible, which allows an upgrade to post-quantum 
cryptographic algorithms once such algorithms
are widely accepted and standardized. Note that certain
solutions need to be redesigned, such as the Butterfly key
expansion (see Section~\ref{sec:butterfly_key_expansion}) since
it assumes a cryptographic algorithm based on the discrete
logarithm problem. 

\subsubsection{Privacy by Design}
A key goal of the security system is to protect the privacy of end
users, in particular people driving vehicles for private use. 
Since most privately owned vehicles are associated with a single 
registered owner, the ability to track the vehicle may be used to 
associate vehicle operation with the registered owner, and so
the system is designed to make it hard to track that vehicle based
on its data transmissions. This is achieved in two ways:

\begin{enumerate}
\item For future applications that involve sending unicast or groupcast (as opposed to broadcast) messages, the application design should use encryption and other mechanisms to prevent the identity of parties to the communication being leaked. Additionally, those applications should be opt-in so that participants have the opportunity to assess the trade-off between privacy and the value of the service offered and only use the application if it offers a good trade-off.\footnote{In practice, it is not clear that end-users are skilled in evaluating these tradeoffs, but this approach is widely accepted in principle.} These privacy-preserving technologies can be implemented without having significant design implications for the SCMS.
\item For applications that involve sending broadcast messages from end-user vehicles, the privacy-preserving aspects of the SCMS defined below are used to make it difficult for eavesdroppers in two physically distant locations to
tell whether BSMs transmitted at the two locations originated from the
same vehicle.
\end{enumerate}

We identify two types of attackers, inside attackers and outside attackers. An outside attacker has access to BSMs but not to any other information such as certificates that have not been broadcast yet. An inside attacker has access to BSMs and to other information, such as information generated during the certificate issuance process. To maintain privacy against outside attackers, we propose
that end-entity devices are issued with a large number of certificates
(we quantify this in section~\ref{sec:cert_provisioning_model}) and
that they make frequent changes in the certificates accompanying BSMs
(e.g.,\ every 5 minutes, or as specified in \cite{j2945.1} -- note that the SCMS supports a range of strategies and that the change strategy is ongoing research from a privacy perspective). To provide defense against inside attackers,
the SCMS operations are divided among its components, and
those components are required to have organizational separation
between them (i.e.,\ each component is run by a separate organization
such that information sharing between organizations can be controlled).
The SCMS is designed such that at least two of
its components need to collude to gain meaningful information for
tracking a device. 

We define ``unlinkability'' informally (cf., Section 4 in \cite{PfitzmannK00} for a more formal definition) as the concept that the greater the distance in time and space between two transmissions from the same device, the harder it is to determine that those two transmissions did in fact come from the same device.

Unlinkability is not a binary property of the system. For example, an eavesdropper who is able to record all messages sent by a vehicle will be able to track that vehicle by constructing the path indicated by that vehicle's BSMs. However, it is a design goal that the V2V communications system does not increase the risk that an individual may be tracked. 

For purposes of the SCMS described in this paper, the requirement is that if a vehicle's broadcast messages (a) contain data that is unique to the vehicle and (b) can be linked to a location, the data should change frequently so that it is extremely difficult for an eavesdropper to track that vehicle.

Note also that any mobile device may potentially be
tracked in other ways that are not addressed by the SCMS (or by any application or data-level mechanism) and are
out of scope of this research, e.g.,\ by RF
fingerprinting~\cite{RFFingerprinting} or by deploying a large-scale
sniffer network. Additional technical mechanisms and possibly legal
regulations are required to counter such attacks. 
%
%

\subsubsection{Misbehavior Detection \& Revocation}
The SCMS enforces revocation by periodically distributing a Certificate
Revocation List (CRL) with an updated list of entries. Devices use the CRL
to identify and reject messages from
revoked devices. In addition, the SCMS maintains {\em internal}
blacklists of revoked devices to deny future certificate requests by them.
For V2V safety applications, each device receives a multitude
of certificates so that traditional CRLs would not work in our
context since they would grow too large. We make the revocation of
devices efficient by using a novel concept of {\em linkage values}
(cf.,~Section~\ref{sec:use_case_pseudo_cert}), which is an extension of
similar ideas presented in \cite{hhl11}. For other applications,
traditional CRLs are supported as well. It is foreseen that CRL
entries will contain some indicator of priority or severity so that recipients who
are limited in storage capacity can store only the $k$ entries of greatest
interest to them, with the minimum value of $k$ system-wide set to 10,000.
%
%
Section~\ref{sec:crl_size_and_distribution} discusses
the CRL size and presents ideas for its dissemination.

Misbehavior detection is the name we give the process of identifying 
devices that are either misbehaving or malfunctioning.
It requires {\em local} misbehavior detection in vehicles to recognize
anomalies, misbehavior reporting by devices to the SCMS, and
{\em global} misbehavior detection by the SCMS to analyze misbehavior
reports and to decide which devices to revoke. The local and global
misbehavior detection algorithms are the focus of on-going work and
are not further discussed in this paper.

\subsection{SCMS Structure}
\label{subsec:full_scms_structure}

In this section we briefly describe the different SCMS components. 
Figure~\ref{fig:overview} shows components within the
system by their logical roles. An implementation of the system may
combine multiple logical roles within a single organization with
proper separation of the logical roles. Components marked V/I provide
separate V2V and V2I functionality while components marked 'X' provide
general functionality for the whole V2X system. Figure~\ref{fig:overview} shows three
pairs of RSEs and OBEs. These are of the same type and used to
illustrate different use cases of the SCMS. The leftmost pair is used
to demonstrate the connections required for bootstrapping, the pair in
the middle shows the connections required for certificate
provisioning and misbehavior reporting, and the rightmost pair shows
the connections required for retrieval of the CRL via the CRL
Store. Connections are marked to show if they have to go through the
Location Obscurer Proxy (LOP), which is an anonymizer proxy and is
defined in detail below. In principle, a single LOP is sufficient to
handle all these types of connections. The content sent over the
connections originating from and ending at the rightmost pair has a
special property. Even if a device is involved only in V2V activities,
it may report misbehavior and receive revocation information on
devices involved in V2I activities (such as an infrastructure
component) and vice versa.  There are two more special types of
connections:

\begin{itemize}
\item SCMS communication: secure communication between components or components and EEs
  supported by the SCMS.
\item Credentials Chain of Trust: this line shows the chain of trust for signature verification. Enrollment certificates are verified against the ECA certificate, pseudonym, application, and identification certificates are verified against the PCA certificate, and certificate revocation lists are verified against the CRL Generator (part of the MA) certificate.
\item Out-of-Band communications: this connection uses a special
  channel and will be explained in more detail in
  Section~\ref{sec:use_case_bootstrap}.
\end{itemize}

All online components communicate with each other using a protected and
reliable communication channel, utilizing protocols such as those from
the Transport Layer Security (TLS) suite~\cite{tls}. Some components
(e.g., Root CA, Electors) are air-gap protected. If data is
forwarded via an SCMS component that is not intended to see that data,
it is further encrypted and authenticated at the application layer
(e.g.,\ to encrypt information generated by the linkage authority that
is addressing the Pseudonym CA but routed via Registration
Authority). The following components are part of the SCMS design:
\begin{itemize}
\item SCMS Manager: Ensures efficient and fair operation of the SCMS, defines organizational and technical policies, sets guidelines for reviewing misbehavior and revocation requests to ensure that they are correct and fair according to procedures.
\item Certification Services: Specifies the certification process and provides information on which types of devices are certified to receive digital certificates.
\item CRL Store (CRLS): A simple pass-through component that stores and distributes CRLs.
\item CRL Broadcast (CRLB): A simple pass-through component that broadcasts the current CRL through RSEs, satellite radio system, etc.
\item Device: An end-entity (EE) unit that sends and/or receives BSMs, e.g., an OBE, an after-market safety device (ASD), or an RSE.
\item Device Configuration Manager (DCM): Used to attest to the Enrollment CA (ECA) that a device is eligible to receive enrollment certificates, and used as a buffer to provide all relevant configuration settings and certificates during bootstrapping.
\item Electors: Electors represent the center of trust of the SCMS. Electors sign ballots that either endorse or revoke an RCA or another elector. Ballots are distributed to all SCMS components, including EEs, to establish trust relationships in RCAs and electors. An elector is established with a self-signed certificate, and the initial set of electors will be trusted implicitly by all entities of the system.
\item Enrollment CA (ECA): Issues enrollment certificates, which act as a passport for the device and can be used to request pseudonym certificates. Different ECAs may issue enrollment certificates for different geographic regions, manufacturers, or device types.
\item Intermediate CA (ICA): This component serves as a secondary Certificate Authority to shield the root CA from traffic and attacks.  The Intermediate CA cert is issued by the root CA.
\item Linkage Authority (LA): Generates pre-linkage values, which are used to form linkage values that go in the certificates and support efficient revocation. There are two LAs in the SCMS, referred to as LA$_1$ and LA$_2$. The splitting prevents the operator of an LA from linking certificates belonging to a particular device. 
\item Location Obscurer Proxy (LOP): Hides the location of the requesting device by changing source addresses, and thus, prevents linking of network addresses to locations.
\item Misbehavior Authority (MA): Processes misbehavior reports to identify potential misbehavior or malfunctioning by devices, and, if necessary, revokes and adds them to the CRL. It also initiates the process of linking a certificate identifier to the corresponding enrollment certificates and adding them to the RA's internal blacklist. The MA contains two subcomponents: Global Misbehavior Detection, which determines which devices are misbehaving; and CRL Generator (CRLG), which generates, digitally signs and releases the CRL to the outside world.
\item Policy Generator (PG): Maintains and signs updates of the Global Policy File (GPF), which contains global configuration information, and the Global Certificate Chain File (GCCF), which contains all trust chains of the SCMS.
\item Pseudonym CA (PCA): Issues short-term pseudonym, identification, and application certificates to devices. Individual PCAs may, for example, be limited to a particular geographic region, a particular manufacturer, or a type of device.
\item Registration Authority (RA): Validates and processes requests from the device. From those it creates singular requests for pseudonym certificates to the PCA. The RA implements mechanisms to ensure that revoked devices are not issued new pseudonym certificates, and that devices are not issued more than one set of certificates for a given time period. Also the RA provides authenticated information about SCMS configuration changes to devices, which may include a component changing its network address or certificate, or relaying policy decisions issued by the SCMS Manager. Additionally, when sending pseudonym certificate signing requests to the PCA or forwarding information to  MA, the RA shuffles the requests/reports to prevent the PCA from taking the sequence of requests as an indication for which certificates may belong to the same batch and the MA from determining the reporters' routes.
\item Root Certificate Authority (RCA): An RCA is the root at the top of a certificate chain in the SCMS and hence a trust anchor. It issues certificates for ICAs as well as SCMS components like PG and MA. The RCA is deployed with a self-signed certificate and trust in a RCA is established by a quorum vote of the electors. A ballot message that contains instructions to add or remove a RCA certificate must be signed by a quorum of the electors. The ballot contains the RCA certificate and the electors' vote. RCA certificates must be stored in secure storage that is usually referred to as a Trust Store. An entity verifies any certificate by verifying all certificates along the chain from the certificate at hand to the trusted RCA. This concept is called chain-validation of certificates, and is the fundamental concept of any PKI. If the RCA and its private key is not secure, then the system is potentially compromised. Due to its importance, an RCA is typically off-line when not in active use.
%
%
%
%
%
%
%
\end{itemize}
Note that the MA, PG, and the SCMS Manager are the only intrinsically-central components of the SCMS.

\subsection{Certificate Provisioning Model} \label{sec:cert_provisioning_model}

The focus in this section is on the provisioning of pseudonym
certificates to OBEs. The provisioning of other certificate types are
either straight-forward, or represent subsets of the pseudonym
certificate provisioning process.  We have developed a pseudonym
certificate provisioning model that balances several conflicting
requirements:
\begin{itemize}
\item \textbf{Privacy vs.\ Size vs.\ Connectivity}: Certificates
  should be used only for short periods of time for privacy
  reasons. The devices cannot store a very large number of
  certificates due to limited memory storage and its cost in a vehicle
  environment. On the other hand, most vehicles cannot establish frequent
  connectivity to the SCMS to download new certificates on demand.
\item \textbf{CRL Size and Retrospective Unlinkability}: The SCMS
  should be able to revoke misbehaving or malfunctioning
  devices\footnote{CRLs can be avoided if devices were required to
  request new certificates frequently such that misbehaving devices
  could be refused new certificates. This would require frequent
  connectivity to the SCMS, which may not be a reasonable
  assumption at least under current circumstances.},
  but putting all valid certificates of a device on
  the CRL would make it very large. We have designed a mechanism to
  revoke a large number of certificates efficiently without revealing
  certificates that were used by the device before it started
  misbehaving.
\item \textbf{Certificate Waste vs.\ Sybil Attack}: Certificates must
  be changed periodically for privacy reasons. One option is to have a
  large number of certificates, each valid one after the other for a
  short period of time. This would result in a large number of unused
  certificates\footnote{Based on the US\ Census Bureau's annual
    American Community Survey for 2011, the average daily commute time
    is less than 1 hour, so more than 95\% of certificates would be
    wasted, if each certificate had a unique validity period.}.
  Another option is to have multiple certificates valid
  simultaneously for longer time periods. This would enable
  masquerading as multiple devices by compromising a single device
  (the so-called \emph{Sybil attack} \cite{sybil}) \cite{PetitSFK15}.
\end{itemize}

In the SCMS model used for Safety Pilot Model Deployment
\cite{safety-pilot}, a certificate was valid for a specific $5$-minute
period, which would correspond to $105,120$ certificates per year.
Given the connectivity constraints, in full deployment a device may need
up to $3$ years' worth of certificates, which would amount
to more than $300,000$ certificates. This approach is
prohibitively expensive in terms of automotive-grade storage
requirements on the device. We studied different variants of this
model, but none of them offer a good balance among all the properties
listed above. Instead, we recommend to adopt a model originating with the CAR 2 CAR
Communication Consortium (C2C-CC) \cite{c2c11}, with certain
modifications to suit our requirements.

In the C2C-CC model, multiple certificates are valid in a given time
period, the certificate validity period is days rather than minutes,
and the certificate usage pattern can vary from device to device,
e.g.,\ a device could use a certificate for $5$ minutes after
start-up, then switch to another certificate, and use that for a
longer time-period before changing certificates again, or even until
the end of the journey. This model offers enough flexibility to find a
good balance among our different requirements except ``CRL size and
retrospective unlinkability'', which we address by our construct of
linkage values (cf.~Section~\ref{sec:use_case_pseudo_cert}). In
particular, our proposal is to use a model originating with the C2C-CC with the following
parameter values:
\begin{itemize}
\item Certificate validity time period: $1$ week
\item Number of certificates valid simultaneously (batch size): minimum $20$
%
%
%
%
\item Overall covered time-span: $1 - 3$ years
\end{itemize}

These parameters can be further refined by the SCMS Manager so that
devices are compatible with each other. Two points are worth noting:
\begin{enumerate}
\item This model provides a reasonable level of privacy against
  tracking while keeping the storage requirements low due to a
  significantly higher utilization of certificates (i.e.,\ far fewer
  certificates are wasted compared to the Safety Pilot
  model). Certificates of a device that does not use all its
  certificates in a week (e.g.,\ say a device only uses 13 of the
  provided 20 certificates in a given week) cannot be
  linked. Moreover, if a device re-uses certificates, it is linkable
  only within a week.

\item The model allows for an easy topping-off mechanism of pseudonym
  certificates. We have designed a mechanism such that devices do not
  need to explicitly request new certificates but the SCMS will
  automatically issue new certificates throughout the life-time of the
  device, until the device stops picking-up certificates for an
  extended period of time. The RA will provide the certificates
  in batches worth one week (e.g.,\ as a zip file), and devices will
  download the batches using TCP/IP connection. The batches will be
  organized in files, named using device information and time
  period. The RA will provide information when to expect new
  certificate batches (e.g.,\ once per month), and the device will
  have full control over what and how much they download, when they
  download, and they can also resume interrupted downloads.
\end{enumerate}

%
%

\section{Certificate Types}
\label{sec:certificate_types}
Within the overall Connected Vehicle system there are different application types which may have different requirements for certificate management. A complete specification of the SCMS therefore should include determining how many different certificate management process flows need to be supported. We define certificates to be of the same \textit{certificate type} if they are issued following identical process flows. Based on an analysis of 119 DSRC-based applications, it was determined that five end-entity certificate types satisfied all use cases:

\begin{itemize}
\item OBE enrollment certificates: enrollment
  certificates are provided during bootstrap and they are used later
  to request message signing and/or encryption certificates.
\item RSE enrollment certificates: RSE enrollment
  certificates are the equivalent to OBE enrollment certificates for
  RSEs.
\item OBE pseudonym certificates: pseudonym certificates provide
  pseudonymity, unlinkability, and efficient revocation of a multitude
  of certificates. This is accomplished by using security features
  such as shuffling, linkage values, butterfly key expansion, and
  encryption of certificates by PCA to the OBE in order to provide
  protection against an insider attack at the RA. This type of
  certificate is used to sign basic safety message (BSM) broadcasts by
  OBEs. Further, this type of certificate is used for authorization
  purposes when unlinkability is required. In the case of BSM
%
%
  broadcast, pseudonym certificates might be attached to each signed
  BSM to increase the ability of a receiver to verify each received
  BSM, or they might be attached to a few signed BSMs
  only~\cite{kw11}.
\item OBE identification certificates: identification certificates are
  used when a device needs to identify itself. This type of
  certificate does not provide pseudonymity nor unlinkability. During
  creation of OBE pseudonym certificates, the SCMS applies privacy
  preserving mechanisms such as shuffling and encryption by the PCA to
  the OBE. These mechanisms are not necessarily used when creating OBE
  identification certificates. However, butterfly key expansion is
  used to allow for continuous generation of
  certificates. Identification certificates are used for authorization
  purposes, such as signed authorization messages of an OBE to an RSE
%
%
  to gain access to a barred road.
\item RSE application certificates: RSEs use these
%
%
  certificates to sign broadcast messages, to sign service
  announcement messages, and optionally to provide an encryption key that
  an OBE can use to send encrypted data. Note that this is the only
  EE certificate type which includes an encryption key.
\end{itemize}

The types of certificates provided by the SCMS are listed in
Table~\ref{table:certificate_types}, along with a list of
features. The table shows required features of each certificate type,
where an 'X' means that a certificate type needs to implement that
feature. The features are as follows:

\begin{itemize}
\item Shuffle: Shuffling is performed in the RA and is useful only in
combination with Butterfly keys that are described below. Shuffling
  makes sure that PCA cannot learn which certificates are assigned to
  which devices. Shuffling is required to provide privacy against SCMS
  insiders.
\item On broadcast CRL if revoked: certificate information of this
  certificate type can be added to a CRL that allows easy recognition
  of revoked certificates. Enrollment certificates and OBE
  identification certificates are never distributed via broadcast CRL
  but they are blacklisted at the RA so that they cannot get new certificates.
\item Simultaneous validity for given PSID: PSIDs~\cite{1609.2} are
  embedded in certificates and the message payload. They describe
  application types and enable the receiver to parse a received
  message. For OBE pseudonym certificates, we allow that more than one
  certificate is valid per time period, as described in
  Section~\ref{sec:cert_provisioning_model} (e.g., 20 certificates are
  valid per week). 
\item Linkage Values: Linkage values are embedded in certificates and
  they allow for efficient revocation,
  cf. Section~\ref{sec:linkage_values}.
\item Continuous Generation: The SCMS will continuously generate
  certificates once an initial request has been made. This feature
  allows for devices to quickly obtain certificates once a connection
  to the SCMS is available and facilitates top-up strategies. This
  feature is particularly useful if a multitude of certificates is
  required.
\item Issuing certificates for multiple time periods: The SCMS will in
  one session issue certificates that are valid for more than one time
  period. For instance, the SCMS might issue certificates each valid
  for one week, but altogether the certificates cover a period of
  several years.
\item Pseudonymity: A pseudonym certificate does not contain any
%
%
  real-world identifier.
\item Misbehavior reporting: Devices provide misbehavior reports to
  the SCMS in order to enable the SCMS to detect misbehavior and
  revoke misbehaving devices. All certificate types that are
  broadcast over-the-air can be included in a misbehavior report.
\item Non-traceability: A device receives multiple certificates such
  that it can use different identities at different times. This raises complexity to
  trace a device, i.e., to determine whether two broadcast messages originate from the same
  device based on the embedded cryptographic material.
\item Encryption key: Besides the public key for signing, the
  certificate holds a second public key to encrypt messages to the
  certificate holder.
\end{itemize}

Table~\ref{table:certificate_types} shows all types of certificates
along with the features they provide. Note the following remarks.

\begin{itemize}
\item Due to the large amount of OBE pseudonym certificates per
  device, linkage values are mandatory for this type. Linkage values
  are not used in OBE identification certificates since those
  certificates are only used for V2I applications and since RSE can
  store large CRLs.
\item We suggest the non-traceability requirement for OBE
  pseudonym certificates only.
\end{itemize}

\begin{table*}[ht]
\caption{Certificate Types}
\centering
\begin{singlespacing}
{\small{
\begin{tabular}{|l|l|l||c|c|c|c|c|c|c|c|c|c|}
\hline
\multicolumn{3}{|c||}{}&
\rotatebox[origin=c]{90}{\,Shuffle\,}&
\rotatebox[origin=c]{90}{\,\parbox{3cm}{
    \begin{center}On broadcast CRL if revoked\end{center}}\,}&
\rotatebox[origin=c]{90}{\,\parbox{3cm}{
    \begin{center}Simultaneous Validity\\ for given PSID\end{center}}\,}&
\rotatebox[origin=c]{90}{\,\parbox{3cm}{
    \begin{center}Linkage Values\end{center}}\,}&
\rotatebox[origin=c]{90}{\,\parbox{3cm}{
    \begin{center}Continuous Generation\end{center}}\,}&
\rotatebox[origin=c]{90}{\,\parbox{3cm}{
    \begin{center}Issuing certificates
      for multiple time periods\end{center}}\,}&
\rotatebox[origin=c]{90}{\,\parbox{3cm}{
    \begin{center}Pseudonymity\end{center}}\,}&
\rotatebox[origin=c]{90}{\,\parbox{3cm}{
    \begin{center}Misbehavior Reporting\end{center}}\,}&
\rotatebox[origin=c]{90}{\,\parbox{3cm}{
    \begin{center}Non-Traceability\end{center}}\,}&
\rotatebox[origin=c]{90}{\,\parbox{3cm}{
    \begin{center}Encryption\end{center}}\,}
\\ \hline \hline
1 & \multicolumn{2}{|c||}{OBE Enrollment Certificate}
& & & & & & & X & & & \\ \hline
2 & \multicolumn{2}{|c||}{OBE Pseudonym Certificate}
& X & X & X & X & X & X & X & X & X & \\ \hline
3 & \multicolumn{2}{|c||}{OBE Identification Certificate}
& & & & & X & X & & X & & \\ \hline
4 & \multicolumn{2}{|c||}{RSE Enrollment Certificate}
& & & & & & & & & & \\ \hline
5a & \multirow{2}{*}{\parbox{2.2cm}{
    RSE Enc.\ \& Appl.\ Certificate}} & Online RSE
& & X & & & & & & X & & X \\ \cline{1-1}\cline{3-13}
5b & & Offline RSE
& & X & & & & & & X & & \\ \hline
\end{tabular}
}}
\end{singlespacing}
\label{table:certificate_types}
\end{table*}

%
%

\section{Butterfly Key Expansion}
\label{sec:butterfly_key_expansion}

The typical process for a device to request certificates from a PKI would be
to generate a private/public key pair. The device creates a certificate signing request (CSR)
including the public key and provide the CSR to the PKI over a secure channel. 
The PKI's CA will then sign the certificate and provide it to the requester.
%
%
For OBE pseudonym certificates, such an approach has disadvantages as
thousands of public keys would need to be generated in the device and
then sent to the SCMS. Butterfly keys are a novel cryptographic
%
%
construction to overcome this disadvantage by allowing an OBE to
request an arbitrary number of certificates; each certificate with a different
signing key and each encrypted with a different encryption key. This is done using
a request that contains only one signing public key seed, one
encryption public key seed, and two expansion functions.  Certificates are encrypted by the PCA to the OBE to avoid the RA being able to relate certificate content with a certain OBE.
Without butterfly keys, the OBE would have to send a unique signing key and
a unique encryption key for each certificate. Butterfly keys reduce
%
%
upload size, allowing requests to be made when there is only
suboptimal connectivity, and also reduce the work to be done by the
requester to calculate the keys. Further, butterfly keys significantly
simplify the topping-off mechanism described in
Section~\ref{sec:cert_provisioning_model}. Butterfly expansion for signing key is
described below for elliptic curve cryptography, but it could easily
be adapted to any discrete-logarithm-type hardness assumption.  Butterfly expansion for encryption key is identical to that of the signing key, except for a minor difference in the way the inputs to AES are  derived, see below for more details.
In the following, we denote integers by lower-case
characters and curve points by upper-case characters. The elliptic
curve discrete logarithm problem is basically the statement: Given $P$
and $A = a P$, but not $a$, it is hard to compute the value of
$a$~\cite{ieee1363}. Butterfly keys make use of this as follows. There
is an agreed base point, called $G$, of some order $l$. The
\emph{caterpillar keypair} is an integer, $a$, and a point $A =
aG$. The certificate requester provides to RA the value $A$ and an
expansion function, $\mathrm{f}_k(\iota)$, which is a pseudo-random
permutation in the integers mod $l$. Note that $\iota$ is a simple
counter iterated by the RA.

In the current design the expansion function for signing keys, $\mathrm{f}_k(\iota)$, which is used to generate points on the NIST curve NISTp256~\cite{NistCurves}, is defined as
\begin{equation}
\mathrm{f}_k(\iota) = \mathrm{f}_k^{int}(\iota) \;\mathrm{mod}\;l , \; \mathrm{where}
\end{equation}
\begin{enumerate}
\item $\mathrm{f}_k^{int}(\iota)$ is the big-endian integer representation of 
\begin{equation}
\begin{split}
 (\mathrm{AES}_k(x+1) \oplus (x+1)) \; \| \; \\
 (\mathrm{AES}_k(x+2) \oplus (x+2)) \; \| \; \\ 
 (\mathrm{AES}_k(x+3) \oplus (x+3)),
\end{split}
\end{equation}
\item  $x+1$, $x+2$, and $x+3$ are obtained by simply incrementing $x$ by 1 each time, e.g., if $x = 01 \ldots 00$, then $x+1 = 01 \ldots 01$, $x+2 = 01 \ldots 10$, $x+3 = 01 \ldots 11$,

\item 128-bit input $x$ for AES is derived from time period $\iota = (i, j)$ as: $\left(0^{32} \| i \| j \| 0^{32} \right)$. 
\end{enumerate}
The expansion function for encryption keys is also defined  as above except $x$ is derived as: $\left( 1^{32} \| i \| j \| 0^{32} \right)$.

Note that in the above definition, AES is used in the Davies-Meyer\footnote{In the Davies-Meyer mode, the output of the function is XORed with the input to generate the final output.} mode,  as $\mathrm{f}_k$ does not need to be invertible. Also, AES is applied 3 times to ensure that the outputs of $\mathrm{f}_k$ are uniformly distributed with negligible biases, if any.  

Now RA can generate up to $2^{128}$ \emph{cocoon public keys} as
$B_{\iota} = A + \mathrm{f}_k(\iota) * G$, where the corresponding
private keys will be $b_{\iota} = a +\mathrm{f}_k(\iota)$, so the
public keys are known to the RA but the private keys are known only to
the OBE. The RA includes the cocoon public keys in the certificate
requests sent to the PCA.

\begin{figure*}
\begin{center}
\scalebox{0.75}{
\input{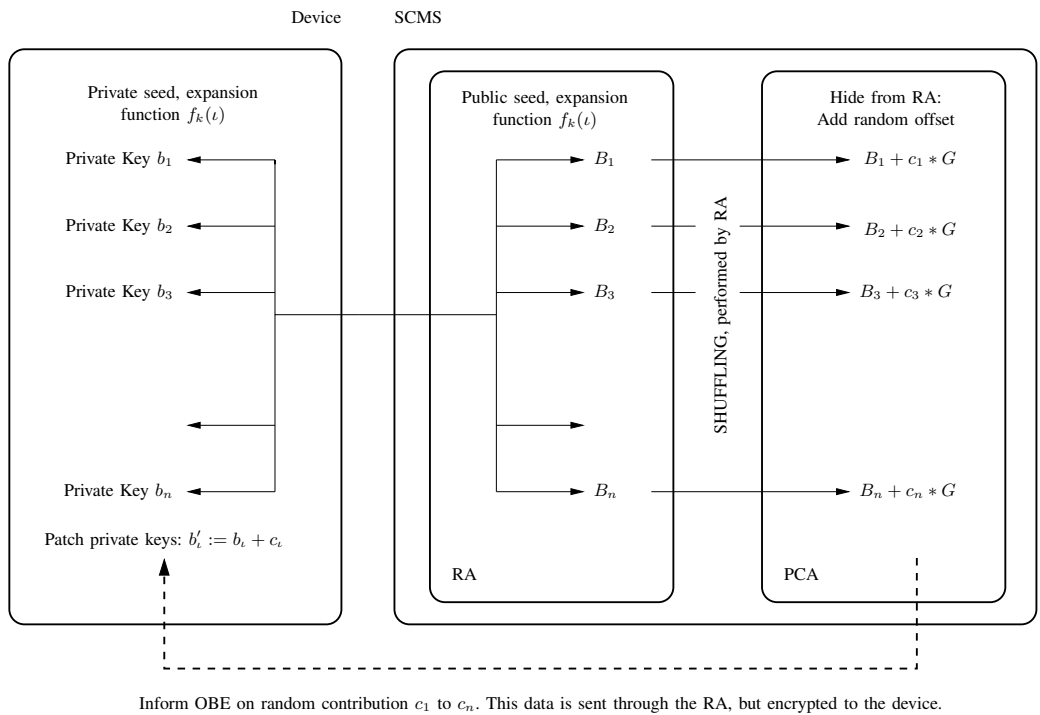}
}
\caption{\label{fig:butterfly_keys}Butterfly Key Expansion Concept}
\end{center}
\end{figure*}

If these expanded public keys were used unaltered by the PCA, the RA,
which knows which public keys come from a single request, could
recognize those public keys in the certificates and track the OBE.
To avoid this, for each cocoon public key $B_{\iota}$, the PCA
generates a random $c_{\iota}$ and obtains $C_{\iota} =
c_{\iota}G$. The 
\emph{butterfly public key} which is included in the
certificate is $B_{\iota} + C_{\iota}$. The PCA returns both the certificate
and the private key reconstruction value $c$ to the RA to be returned
to the OBE\footnote{This description covers explicit certificates,
i.e., certificates that include the public key explicitly. In fact,
the SCMS as currently implemented issues 
%
%
\emph{implicit certificates}~\cite{implicit11} for devices. The implicit certificate
generation process inherently changes the public key in a way that
leaves input and output uncorrelatable by the RA and is consistent
with the motivation for butterfly keys.}. To prevent the RA from
working out which certificate corresponds to a given public key in a
request, the certificate and the reconstruction value $c_{\iota}$ are
encrypted to the OBE. The OBE will update its private keys resulting in
$b'_{\iota}$ with $b'_{\iota} = b_{\iota} + c_{\iota}$. To prevent the
PCA from knowing which certificates go to which OBE, each
certificate must be encrypted with a different key. The encryption
keys are also generated with the butterfly key approach: the OBE
provides a caterpillar encryption public key $H = hG$, the RA expands
it to cocoon public encryption keys $J_{\iota} = H + f_e(\iota)G$, and
the PCA uses these keys to encrypt the
response. Figure~\ref{fig:butterfly_keys} provides an overview of the
butterfly key expansion concept for signing keys.

\subsection{Security of Butterfly Keys}
In this section, we briefly discuss the security of butterfly keys. We say that the above construction of butterfly keys is secure, if any efficient (i.e., polynomial-time) adversary has only a negligible (i.e., smaller than any polynomial) chance of winning in the following game: adversary receives a polynomial number of butterfly public keys (i.e., $(A + \mathrm{f}_k(1) * G + c_1 * G), (A + \mathrm{f}_k(2) * G + c_2 * G), ..., (A + \mathrm{f}_k(q) * G + c_q * G)$), and it needs to figure out at least one of the butterfly private keys (i.e., $a +\mathrm{f}_k(x) + c_x$ for any $1 \leq x \leq q$). As $\mathrm{f}_k$ is assumed to be a pseudo-random permutation, it is hard for any polynomial-time adversary to distinguish $\mathrm{f}_k$'s outputs from truly random values. Hence, we can simplify (without any loss of generality) the above security definition as follows.

\textbf{Security Definition (Informal)}. The above construction of butterfly keys is said to be secure, if any efficient adversary that is given a polynomial number of butterfly public keys (i.e., $(A + b_1 * G), (A + b_2 * G), ..., (A + b_q * G)$, where $b_1, b_2, ..., b_q$ are randomly chosen) has only a negligible probability of correctly guessing one of the butterfly private keys (i.e., $a + b_x$ for any $1 \leq x \leq q$).

\textbf{Theorem (Informal)}. The construction of butterfly keys is secure assuming that the elliptic curve discrete logarithm problem is hard.

\textit{Proof.} We prove the above theorem by contradiction. We first assume that the construction of butterfly keys is {\em not} secure (i.e., there exists a polynomial-time adversary, henceforth butterfly-keys-adversary, with a non-negligible probability of winning the security game), then using this adversary we construct another polynomial-time adversary, henceforth discrete-log-adversary, that solves the elliptic curve discrete logarithm problem with a non-negligible probability, although with a polynomial loss in the winning probability. However, as the theorem says the latter can not be true, the construction of butterfly keys must be secure.

The discrete-log-adversary is given a pair of curve points $(P, A)$, and it needs to output $a$, s.t. $A = aP$. It randomly selects $q$ integers $b_1, b_2, ..., b_q$, and using them generates $q$ butterfly public keys $(A + b_1 * P), (A + b_2 * P), ..., (A + b_q * P)$, and gives them to the butterfly-keys-adversary. When the butterfly-keys-adversary returns its response $c$, the discrete-log-adversary uses $c$ to compute its own response as follows: pick a random number from $1$ through $q$, say $y$, now the response is $c - b_y$.

It is clear that the discrete-log-adversary runs in polynomial-time if the butterfly-keys-adversary runs in polynomial-time. It remains to be shown that the winning probability of discrete-log-adversary is non-negligible. To this end, we note that if the butterfly-keys-adversary wins, then $c = a + b_x$ for some $1 \leq x \leq q$. Since, $y$ is picked at random from $1$ through $q$ after the butterfly-keys-adversary has responded, $y = x$ with probability $1/q$, and hence $c - b_y = a$ with probability $1/q$ times the winning probability of butterfly-keys-adversary. Therefore, if the winning probability of butterfly-keys-adversary is non-negligible, so is that of discrete-log-adversary.~$\blacksquare$

%
%

\section{Device Bootstrapping and Certificate Provisioning}
\label{sec:use_cases}

The SCMS supports four main use cases: device bootstrapping, certificate
provisioning, misbehavior reporting, and global misbehavior detection
\& revocation. Bootstrapping and certificate provisioning will be
presented in this section, and misbehavior reporting, detection and
revocation will be described in the next section.

\subsection{Device Bootstrapping}
\label{sec:use_case_bootstrap}
Bootstrapping is executed at the start of the life cycle of a device. It
equips the device with all the information required to communicate
with the SCMS and with other devices. It is required that correct
information is provided to the device during bootstrapping, and that
the CAs issue certificates only to certified devices. Any bootstrapping
process is acceptable that results in this information being
established securely.

The bootstrapping process includes a device, the DCM, the ECA and the
certification services component. We assume that the DCM has
established communication channels with other SCMS components, such as the ECA or the policy generator, and that it will communicate with the device to be
%
%
bootstrapped using an out-of-band channel in a secure environment. 
Bootstrapping consists of two operations: initialization and enrollment. 
Initialization is the process by which the device obtains certificates it needs 
to be able to trust received messages. Enrollment is the process by which the
device obtains an enrollment certificate that it will need to sign messages
to the SCMS.

Information received in the initialization process includes (1) the certificates 
of all electors, all root CAs, and possibly of intermediate CAs as well as PCAs,
and (2) the certificates of the misbehavior authority, policy generator, and the CRL
generator to send encrypted misbehavior reports and verify received policy files and CRLs.

\begin{figure*}[ht]
\begin{center}
\scalebox{0.75}{
\input{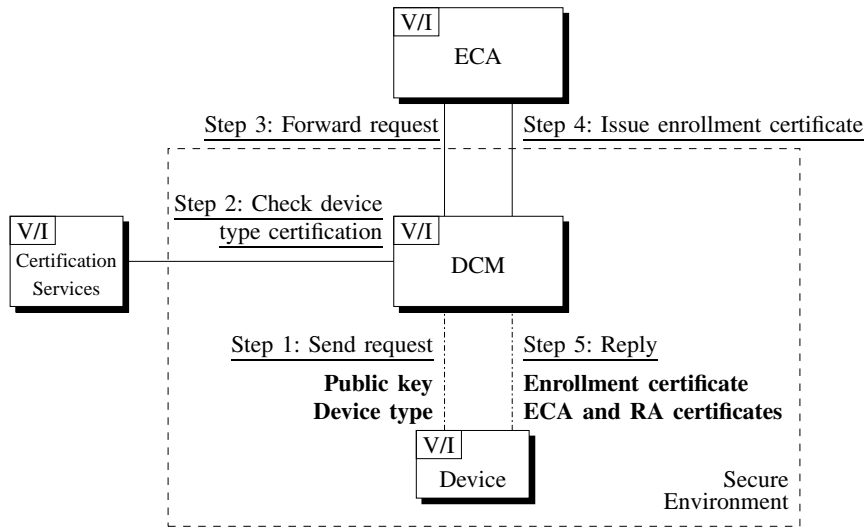}
}
\caption{\label{fig:enrollment}Exemplary Enrollment Process}
\end{center}
\end{figure*}

In the enrollment process, information required to actively
participate is sent to the device. This includes (1) the enrollment
certificate, (2) the certificate of the ECA, and (3) the certificate
of the RA and information necessary to connect to the RA. During the
enrollment process, the certification services provide the DCM with
information about device models which are eligible for
enrollment. This requires that the DCM receives trustworthy
information about the type of device to be enrolled. 
Figure~\ref{fig:enrollment} shows an exemplary enrollment process.

%
%

\subsection{Overview of Certificate Provisioning}
\label{sec:use_case_pseudo_cert}
The certificate provisioning process is most complex for OBE pseudonym
certificates in order to protect end-user privacy and to minimize the
required computational effort on the resource-constrained device. In
the following we focus on the pseudonym certificate provisioning of
OBE pseudonym certificates since the certificate provisioning of other
certificate types is a straight-forward subset in terms of
functionality.  The OBE pseudonym certificate provisioning process is
illustrated in Figure~\ref{fig:certificate_provisioning}, and is
designed to protect privacy from inside and outside
attackers. The SCMS is designed to ensure
that no single component knows or creates a complete set of data that
would enable tracking of a vehicle.  The RA knows the enrollment
certificate of a device that requests pseudonym certificates, but even
though the pseudonym certificates are delivered to the device by the
RA, it is not able to read the content of those certificates; the PCA
creates each individual pseudonym certificate, but it does not know the
recipient of those certificates, nor does it know which certificates
went to the same device. The LAs generate hash-chain values that are
masked and embedded in each certificate, and unmasked by publishing a
secret linkage seed pair to efficiently revoke all future device
%
%
certificates. However, a single LA is not able to track devices by
linking certificates or to revoke a device, but both LAs, the PCA, and
the RA need to collaborate for the revocation process.  Privacy
mechanisms in the SCMS include:
\begin{itemize}
\item \textbf{Obscuring Physical Location.} The LOP obscures the physical location of an end-entity device to hide it from the RA and the MA.

\item \textbf{Hiding Certificates from RA.} The \emph{butterfly key expansion} process ensures that the public key seeds in requests cannot be correlated with the resulting certificates. Details were given in Section~\ref{sec:butterfly_key_expansion}. Encrypting the certificates to the device prevents the RA from relating certificates with a device.

\item \textbf{Hiding Receiver and Certificate Linkage from PCA.} The RA expands incoming requests using Butterfly keys and then splits these requests into requests for single certificates, and shuffles requests of all devices before sending them to the PCA. This prevents the PCA from learning whether two different certificates went to the same device, which would violate our privacy goal by enabling the PCA to link certificates. The RA should have configuration parameters for shuffling, e.g., the POC shuffles either 10,000 requests or a day's worth of requests, whatever is reached first.
%
%
\end{itemize}

\subsection{Linkage Values}
\label{sec:linkage_values}
\begin{center}
\begin{figure*}[ht]
\scalebox{1.05}{
\input{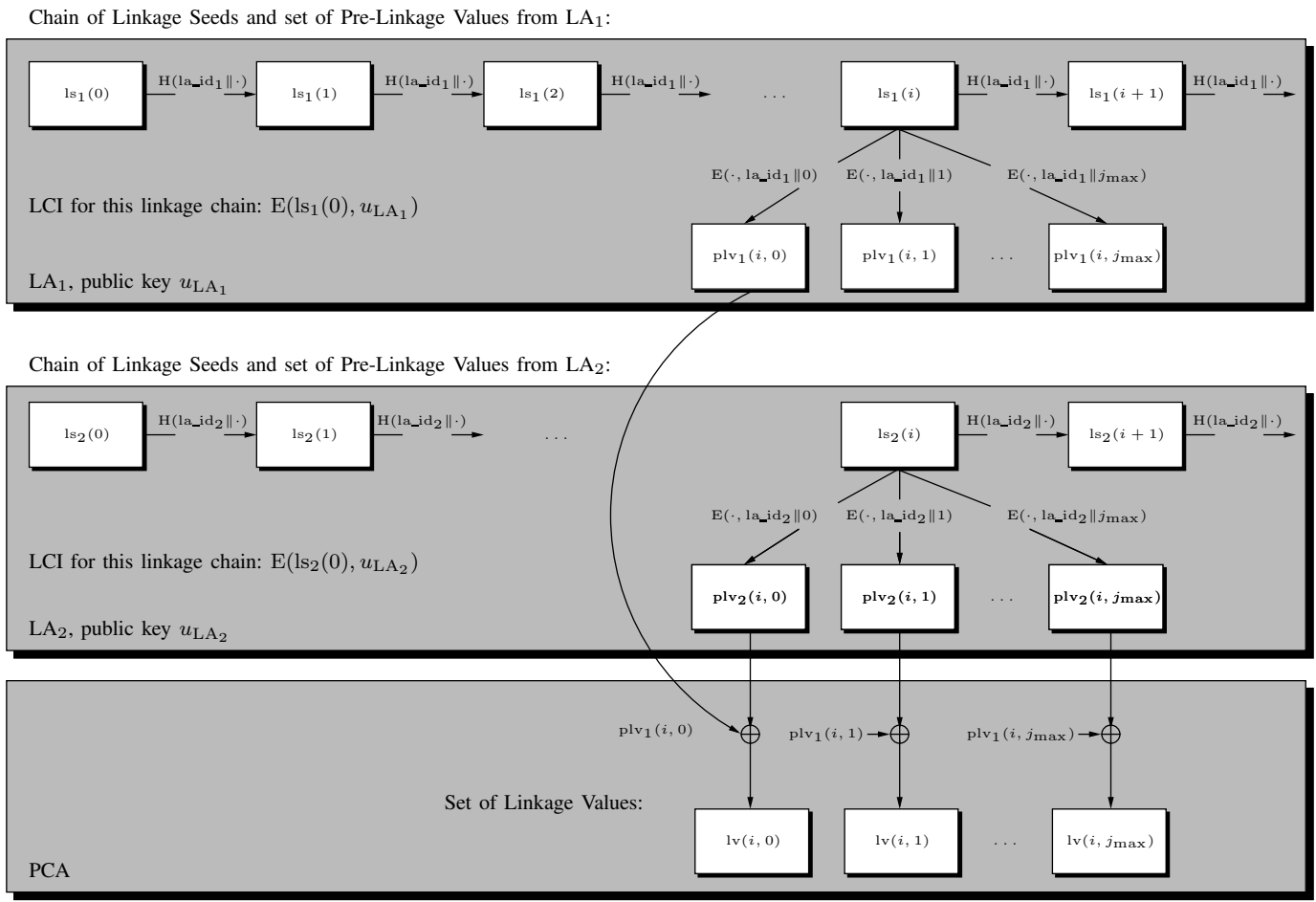}
}
\caption{\label{fig:linkage_value_creation}Creation of Linkage Values}
\end{figure*}
\end{center}

For any set of pseudonym certificates provided to a device, the SCMS
inserts \emph{linkage values} in certificates that can be used to
revoke all of the certificates with validity equal to or later than
some time $i$. These linkage values are computed by XORing the
pre-linkage values generated by the Linkage Authorities LA$_1$ and
LA$_2$. The LAs can generate the pre-linkage values in advance.
Figure~\ref{fig:linkage_value_creation} provides an overview of the linkage value generation.

Let $\laid_1, \laid_2$ be $32$-bit identity strings associated with
  LA$_1$, LA$_2$, respectively. For a set of certificates, first the
  LA$_1$ (resp., the LA$_2$) picks a random $128$-bit string called
  the initial linkage seed $\ls_1(0)$ (resp., $\ls_2(0)$), then for
  each time period (e.g., a week) $i > 0$ calculates the linkage seed
  $\ls_1(i) \leftarrow \H_u\left(\laid_1 \; \| \; \ls_1(i-1)\right)$ (resp., $\ls_2(i) \leftarrow
  \H_u(\laid_2 \; \| \; \ls_2(i-1))$). In this coherence, $\H_u(m)$ denotes
  the $u$ most significant bytes of the SHA-256 hash output on $m$,
  and $a \| b$ denotes concatenation of bit-strings $a$ and $b$. We suggest to use $u=16$. Note that the linkage seeds (i.e., hash chains) created by the LAs have the property that it is easy to  calculate forward (i.e., $\ls(i)$ from $\ls(i-1)$) but it is computationally infeasible to calculate backward (i.e., $\ls(i-1)$ from $\ls(i)$). Now pre-linkage values are calculated by means of a pseudorandom function. We choose to implement this by an encryption function,  such as AES, in the Davies-Meyer mode. Each LA encrypts the linkage seeds as $\plv_x(i, j)
  \leftarrow \left[\E(\ls_x(i), (\laid_x \| j)) \oplus (\laid_x \| j)\right]_v, \;  x \in \{1, 2\}$, where $\E(k, m)$ is the AES encryption of $m$ with key $k$, $a \oplus b$ is the exclusive-OR of bit-strings $a, b$, and $[a]_v$ denotes the $v$ significant bytes of bit-string $a$. We suggest flexible use of $v$ to account for the number of deployed
devices and potential weaknesses of the underlying cryptographic
primitives in terms of collision resistance. Currently, $v=9$
appears to suffice. The value $i$ denotes a time period (e.g., a week) and $j$ denotes certificates within a time period (e.g., 20 certificates per week). Each LA calculates pre-linkage values in the same manner, but each with randomly selected initial seed. We denote the resulting values as $\plv_1$ and $\plv_2$. In order to select a specific linkage chain from an LA, we use Linkage Chain Identifiers (LCI)s. An LCI is 
the initial linkage seed $\ls_1(0)$ or $\ls_2(0)$ that LA$_1$ or LA$_2$, respectively, encrypts
to itself, e.g., $\E(pk_1, \ls_1(0))$, where $pk_1$ is the public key of LA$_1$.

Pre-linkage values are encrypted (individually) for the PCA but sent to the RA for association with a certificate request. The PCA XORs the pre-linkage values to obtain the linkage value $\lv =
  \plv_1 \oplus \plv_2$. Similar processing is required when the CRL is processed by a device. The details of this process and the information that needs to be published are presented in Section~\ref{sec:crl_processing}.

\subsubsection{Hiding Linkage Information}
The PCA computes the linkage value to be included in a certificate by
XORing together the two pre-linkage values from the LAs, which are
generated independently by the two LAs and encrypted for the PCA to
prevent the RA from colluding with one of the LAs and mapping
pre-linkage values to linkage values. Therefore, no single component
is able to link pseudonym certificates of a single device. 

\subsubsection{Response Encryption}
The PCA creates singular certificates which are collated by the RA and made available to the device. To prevent the RA from seeing multiple certificates belonging to the same device, the PCA encrypts each singular certificate to the device. The PCA and the device use the butterfly key expansion process to encrypt each certificate with a different key. 

\subsubsection{Linkage Value Length}
The linkage values and pre-linkage values are chosen to be 9 bytes in length. This length is
%
%
appropriate because the unique identifier is not just linkage value
but ($i$, linkage value); this length ensures negligible probability
that two devices use the same linkage value in the same time period.
Consider that there are $2.5 * 10^8$ cars and each has on average 40
certificates per linkage period (one week), then in any linkage period
identified by a given $i$, there are $10^{10} \approx 2^{33}$ linkage
values in use. The relevant quantity when calculating collision
probabilities is the number of pairs of values, which in this case is
about $2^{66}$. With 72-bit linkage values, this means that the chance
of a collision between two linkage values is about $2^{-6}$. In other
words, there will be about one collision every 64 linkage periods,
which with linkage period length of a week means about one linkage
value collision in the system of 250 million vehicles every year.
This is small and in addition a linkage value collision matters only
in case of a revocation, i.e.,
%
%
if a linkage value was assigned to a revoked device. Assuming revocation rates below 1\%, this
implies that a linkage value collision that actually makes a
difference will happen less than once every 100 years. This could be
reduced further by increasing the size of the linkage value. However,
that would increase the size of all certificates, which would increase
channel congestion.
%
%
%
%
\subsection{Misbinding Attacks}
In this type of attack, an attacker Mallory misuses the public-key of her target Alice. Mallory reads Alice's public key and requests a certificate over that public key from the SCMS. 
While Mallory does not know the corresponding private key, she can still mount attacks in which she gets a message signed by Alice's private key and then attaches her certificate rather than Alice's. In the context of the BSM this is not a particularly significant attack, but SCMS-issued certificates could be used for a wide range of applications and in some of those applications the attack could have a greater impact. It is therefore useful to consider countermeasures.

The countermeasure chosen is the one specified in~\cite{1609.2}: when a message is signed, the hash that is signed is calculated over both the message itself and the certificate. This binds the hash to the certificate and provides assurance that the certificate provided with the message is in fact the certificate that the sender intended to be used. Since this certificate misbinding attack is only of any use if the false certificate is different from the true one, this approach of including the hash completely eliminates this attack.

Other mechanisms were considered. For example, this attack is possible only if the certificate request messages do not provide proof of possession; and this is the case in the SCMS, since operational certificate requests are not signed with the key for the certificate to be issued but with the key for the enrollment certificate. Signing certificate requests with the enrollment certificate is a key part of the trust architecture of this system and so cannot be changed, but we considered signing the certificate request with both the enrollment certificate and the private key for the certificate to be issued. However, this has shortcomings compared to hashing the certificate into the message. First, if a requester were to legitimately request two certificates for the same key pair, misbinding would still be possible. Second, hashing the certificate into the message allows the receiver to determine for sure that the certificate is the one intended, rather than relying on the CA enforcing a particular mechanism for certificate request. For these reasons we determined that proof-of-possession, although it is widely used in internet protocols, does not add significant value in our system and that including the certificate in the message hash is the only countermeasure necessary.

\subsection{Detailed Description of Pseudonym Certificate Provisioning Process}
\begin{center}
\begin{figure*}[ht]
\scalebox{0.93}{
\input{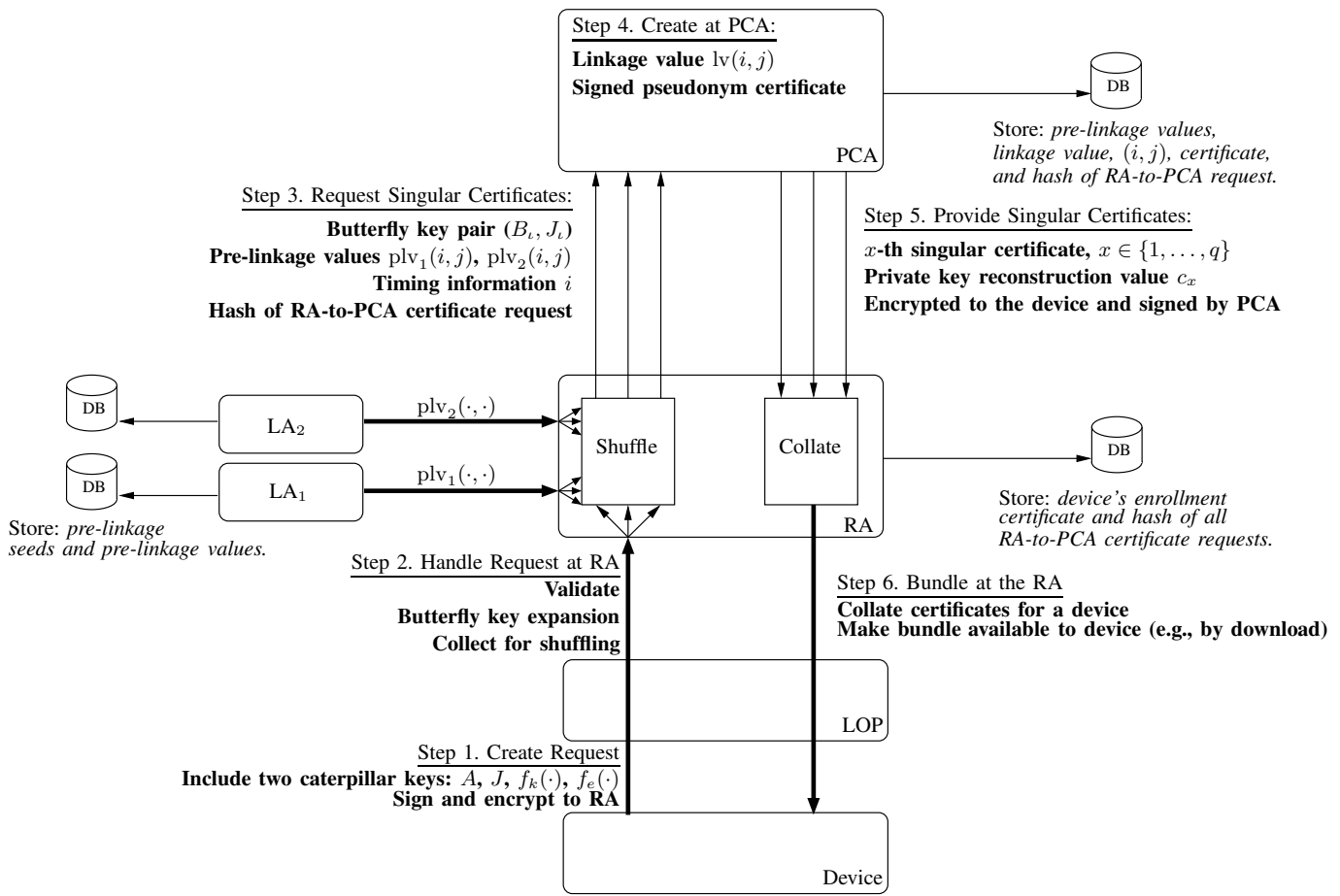}
}
\caption{\label{fig:certificate_provisioning}Certificate Provisioning}
\end{figure*}
\end{center}

In the following, we present a detailed step-by-step description of the pseudonym
certificate provisioning process, illustrated in Figure~\ref{fig:certificate_provisioning}.

\begin{itemize}
\item \textbf{Step 1.} The device creates a request by generating
  butterfly key seeds, signing the request
  with the device's enrollment certificate, and encrypting the request to the RA.  The device then sends the request to the RA via LOP. The LOP
%
%
%
%
%
%
  functions as a pass-through device for requests. The LOP obscures
  the device's identifiers (e.g., IP address) by replacing these
  identifiers with its own in transit, such that the request appears
  to the RA as coming from the LOP. The
  functionality of the LOP is very similar to the masquerading feature
  implemented in many Internet routers.

\item \textbf{Step 2.} The RA decrypts the request, checks
  if the signature and the device's enrollment certificate are valid
%
%
%
%
  and that the latter is not revoked. Further, it checks if this is
  the only request by the device. If
%
%
  all checks succeed, the RA sends an acknowledgment to the device,
  and performs the butterfly key expansion as explained in
  Section~\ref{sec:butterfly_key_expansion}. Otherwise, the request is
  denied. The RA collects several such requests from different devices
  along with the sets of pre-linkage values received from the
  LAs. Once enough such requests are available, the RA shuffles the expanded individual requests
  to the PCA.
%
%

  Note that during pre-generation of additional pseudonym certificates, the RA 
%
%
%
%
  requests pre-linkage values from each of the LAs for a particular
  initial linkage seed that is associated with that device. This can
  be accomplished using the LCI for the corresponding linkage chain. 

\item \textbf{Step 3.} The RA sends requests for pseudonym
  certificates to the PCA, one certificate per request, where each
  request consists of a to-be-signed certificate, a response encryption public key, and an encrypted
 pre-linkage  value from each of the LAs ($\plv_1(i,j), \plv_2(i,j)$) , and the hash of  the RA-to-PCA 
 pseudonym certificate request.

\item \textbf{Step 4.} The PCA decrypts the pre-linkage values and computes the
 linkage value $\lv(i,j) = \plv_1(i,j) \oplus \plv_2(i,j)$. It then adds the
 linkage value to the to-be-signed certificate, and implicitly signs it to create
 a pseudonym certificate and a private key reconstruction value. Subsequently,
 it encrypts both the pseudonym certificate and the private key reconstruction
 value, using the response encryption public key.

\item \textbf{Step 5.}  The PCA signs the encrypted packet generated in step 4
  above, and sends it to the RA. Signing the encrypted packet provides
  a guarantee to the device that the PCA encrypted the packet for the
  device. This prevents a man-in-the-middle attack where an insider at
  the RA substitutes the valid response encryption key with another
  key for which the RA knows the private key, and thus the RA may be able to
  see the contents of the pseudonym certificate including the linkage value.

\item \textbf{Step 6.} The RA collects the encrypted pseudonym certificate and
  corresponding private key reconstruction value for one week and bundles them
  for a given device. These bundles are called batch. The RA provides the batches
  to the device for download.
\end{itemize}

For revocation purposes, the SCMS components involved in the pseudonym
certificate provisioning store different information corresponding to
a given pseudonym certificate as listed in
Table~\ref{table:cert_provision}.

\begin{table}[ht]
\caption{Information Stored by SCMS Components}
\centering
\begin{tabular}{|l|p{6.3cm}|}
\hline
Component & Stored Information\\ [0.5ex]
\hline \hline
LA & Initial linkage seed, pre-linkage value \\ \hline
\multirow{3}{*}{PCA} &
Encrypted pre-linkage values from both LAs and their
corresponding $(i, j)$ values, linkage value,
certificate, and hash of RA-to-PCA pseudonym certificate request\\ \hline
\multirow{2}{*}{RA} &
Enrollment certificate and its validity period,
hash value of RA-to-PCA pseudonym certificate request \\ [1ex]
\hline
\end{tabular}
\label{table:cert_provision}
\end{table}

%
%

\section{Removing Misbehaving Devices}
The removal of misbehaving devices in an efficient manner is an essential design objective. We separate the removal of misbehaving devices into (1) reporting misbehavior, (2) globally detect misbehavior, (3) investigate misbehavior, and (4) revoke a misbehaving EE. 

\subsection{Misbehavior Reporting}
\label{sec:use_case_mb_reporting}
V2V messages from misbehaving or defective devices can contain false
or misleading information. We distinguish between intentional and
unintentional misbehavior, where the latter covers all faults and
error cases of devices. In both cases it is important that benign
participants neglect messages from misbehaving devices. One approach
to accomplish this is to run misbehavior detection algorithms on the
device (local misbehavior detection) in order to identify misbehaving
nodes. A second approach is to report potentially misbehaving devices
to the SCMS. The SCMS will run misbehavior detection algorithms and
then inform all participants about certificates which are no longer
trustworthy. In the misbehavior reporting process, devices will send
misbehavior reports to the MA via the RA . The RA which sits behind the LOP will not know the source IP. The RA will combine and shuffle the reports from multiple reporters to
prevent the MA from tracking the reporter's path based on the
reports. The format of a misbehavior report is not fully defined yet,
but a report will include reported (suspicious and alert-related) BSMs,
associated pseudonym certificates, as well as the reporter's pseudonym certificate
and corresponding signature from the time the report was created 
(compare preliminary EE-MA interface~\cite{scms-ee-ma-interface}). A report 
will be encrypted by the reporter to the MA. In the following, we will 
first focus on the process of Global Misbehavior Detection and Revocation 
for the case of OBE pseudonym certificates. Subsequently, we describe the
processes required for other types of certificates.

\subsection{Global Misbehavior Detection}\label{sec:use_case_revocation_obe_certs}
The global misbehavior detection is the overall process to identify potential misbehavior in the system,
investigate suspicious activity, and if confirmed, to revoke misbehaving EEs. The Misbehavior Detection
process is owned and executed by the MA. A number of Global misbehavior detection algorithms have
been developed as part of another research project. Those will be integrated to the current SCMS design
%
%
and be tested in various CV Pilot activities. However, as the V2X landscape continues to evolve and new
threats and forms of misbehavior are discovered, it's expected that additional algorithms will continue
to be developed and implemented over time. Misbehavior Detection methods and algorithm development
are seen as iterative tasks that will continue throughout the lifetime of the SCMS. One example of 
misbehavior, however primitive, would be a bad actor who intentionally projects the position of the
sending vehicle 3 meters to the left (or right for right hand drive countries). These messages would
cause alerts to the oncoming traffic which would be detected as possible misbehavior. The receiving
vehicle would store these messages (assuming multiple) and put them into a MB report, along with all
defined data and details. This report will be encrypted and sent to the RA for submission to the MA. As
other vehicles also detect this misbehaving vehicle they would also send MB reports to the MA. As the
number of reports grow it would trigger the MB detection algorithms and kick off the misbehavior
investigation process possibly leading to a device revocation.

It's worth noting that this type of misbehavior could possibly be detected locally by the sending vehicle. OEMs, and EE device developers are expected to tackle misbehavior, at the device level, from many angles to detect and prevent bad messages from being sent.
%
%
%
%
%
%
Misbehavior Detection requires that the MA be able to learn whether multiple misbehavior reports point to the same device. In order to revoke a device's certificates, the MA needs to collect information that determines the revocation information to include in a CRL for distribution. Additionally, the MA needs to provide the RA with information required to perform blacklisting. The following actions are required from the components:
\begin{enumerate}
\item The MA, PCA and one of the LAs have to collaborate to reconstruct
  linkage information.
\item The MA, PCA, RA, and both the LAs have to collaborate to produce
  external revocation information for the CRL.
\item The MA, PCA and the RA have to collaborate to determine the
  enrollment certificate of the misbehaving device, which will be added
  to the RA's blacklist.
%
%
\end{enumerate}

Step 1 is executed as part of the Misbehavior Investigation to determine whether an EE, or a set of EEs, did indeed misbehave. After an EE has been marked as misbehaving, Steps 2 and 3 are executed as part of Revocation to determine the information that is included in the CRL and that is added to the internal blacklist. 

\subsection{Misbehavior Investigation}
Misbehavior Investigation is the task to determine whether suspicious activities are indeed due to misbehavior, and to identify misbehaving EEs. This process is initiated by the Misbehavior Detection algorithm running in MA, and it depends on PCA and an LA. This separation introduces a checks and balances into the system, and we strongly recommend a mechanism that PCA and LA require some kind of proof of MA (e.g., a particular authorizing digital signature of each request), and PCA and LA limit the number of requests as well as the amount of information returned to MA. Finally, we strongly recommend that PCA and LA keep records of every request, and that these log files are regularly audited. 

The Misbehavior Investigation process reveals in a controlled manner linkage information of EEs for which two (or more) input certificates were generated for the same EE. In the following, we present a detailed step-by-step description of this process. Note that Steps 1 and 2 are included for completeness and cover Misbehavior Reporting Global Misbehavior Detection, respectively. 
\begin{itemize}
\item \textbf{Step 1.} The MA receives misbehavior reports, including
  a reported pseudonym certificate with linkage value $\lv = \plv_1
  \oplus \plv_2$.

\item \textbf{Step 2.} The MA runs global misbehavior detection algorithms to
  determine which reported pseudonym certificates might be of
  interest, i.e.,\ for which pseudonym certificates linkage information
  needs to be retrieved.

\item \textbf{Step 3.} The MA makes a request (signed) to the PCA to
  map the linkage values $\lv$ of the identified pseudonym
  certificates to the corresponding encrypted pre-linkage values $(\plv_1,
  \plv_2)$ 
  from the PCA's database. The PCA returns the encrypted pre-linkage values to MA.

\item \textbf{Step 4.} The MA makes a request to either LA$_1$
  or LA$_2$ in order to find out whether two instances of encrypted $\plv_1$
  (or, resp., $\plv_2$) point to the same device. Note that for this
  purpose, the LA can respond with binary information, indicating
  whether the two presented encrypted pre-linkage values point to the same
  device. There may be extensions where the MA presents a multitude of
  pre-linkage values and the LA reveals information about their correlation
  in a controlled manner.
\end{itemize}

\subsection{Revocation and Blacklisting.}  
\begin{center}
\begin{figure*}[ht]
\scalebox{0.72}{
\input{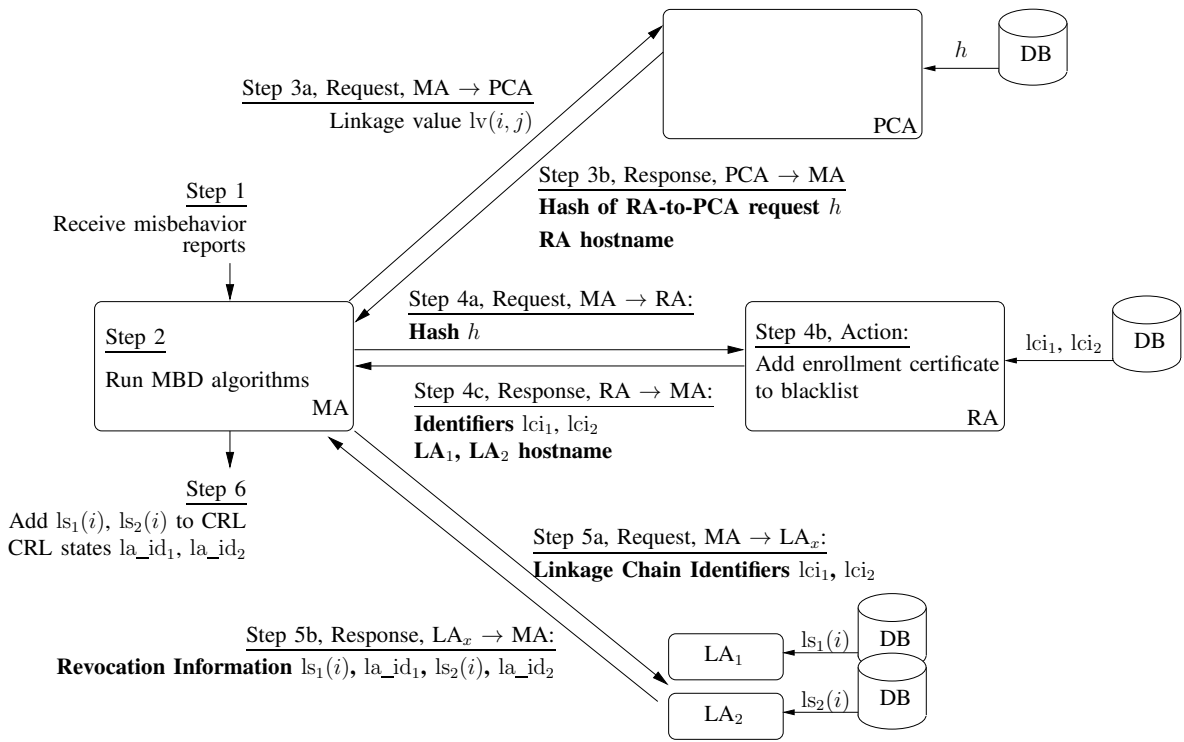}
}
\caption{\label{fig:certificate_revocation}Pseudonym Certificate Revocation}
\end{figure*}
\end{center}

If an EE has been identified as misbehaving during the Misbehavior Investigation, 
the EE is revoked and blacklisted. Next we present a detailed
step-by-step description of the Revocation and
Blacklisting process to identify the linkage seeds and the enrollment
certificate corresponding to a pseudonym certificate. 
Figure~\ref{fig:certificate_revocation} illustrates this process, with Steps 1 and 2
summarizing the Misbehavior Investigation. Note that some of the 
communication messages in the steps below are digitally signed.

%
%
%
%
%
%
\begin{itemize}
\item \textbf{Step 3.} The MA makes a request (signed) to the PCA to
  map the linkage value $\lv$ of the identified pseudonym certificate
  to the corresponding hash value of the RA-to-PCA pseudonym
  certificate request, all from the PCA's database. The PCA returns
  this value and the hostname of the corresponding RA to the MA.

\item \textbf{Step 4.} The MA sends the hash value of the
  RA-to-PCA pseudonym certificate request (signed) to the RA so that
  the RA can add the corresponding enrollment certificate to its 
  blacklist. The RA does not reveal the enrollment certificate though. The MA
  receives the following information from RA, which is then used by MA to revoke EEs
  via CRL entries:

\begin{itemize}
\item The hostnames of the LAs involved in creating the considered
  linkage values,
\item For each LA, an array of LCIs is provided. An LCI can be used by the LA to look up
  the linkage chain and the underlying linkage seed. Multiple linkage chain identifiers
  are only used if a device owns certificates from multiple
  independent linkage chains, which is considered an exception.
\end{itemize}

\item \textbf{Step 5.} The MA makes a request to the LA$_1$
  (resp., the LA$_2$) to map LCI $\lci_1$ (resp., $\lci_2$) to the linkage
  seed $\ls_1(i)$ (resp., $\ls_2(i)$), where $i$ is the currently
  valid time period. Both the LAs return the linkage seed to the MA.
  Further, each LA provides to MA its linkage authority ID
  ($\laid_i$).  Note that given a linkage seed $\ls_1(i)$ and the
  corresponding $\laid_i$, only the {\em forward} linkage seeds (i.e.,
  $\ls_1(j)$ for $j \geq i$) can be calculated, and thus {\em
    backward} privacy of the revoked device is maintained.

\item \textbf{Step 6.} The linkage seeds $\ls_1(i)$ and $\ls_2(i)$,
  are added to the CRL. The CRL globally states the current time
  period $i$. Further, it states the corresponding pair of LA IDs,
  $\laid_1$, $\laid_2$. For efficiency reasons, the CRL may group
  entries with the same LA ID pair together to save over-the-air bytes.  
 Then the MA's CRLG signs the CRL and publishes it.
\end{itemize}

\subsection{Global Misbehavior Detection \& Revocation for
  Non-pseudonym Certificates}
\label{sec:use_case_revocation_other_certs}

The misbehavior detection and revocation process shown above is
tailored to the case of pseudonym certificates. This type of
certificate is the only type using linkage values which allows for
efficient revocation of a multitude of certificates. All other types
of certificates allow for a simpler process for misbehavior detection
and revocation. Without linkage values, all revoked certificates need
to be identified separately on the CRL in order to revoke them. The
process reads as follows:

\begin{itemize}
\item \textbf{Step 1.} The MA receives misbehavior reports, including
  a reported certificate. This certificate does not contain linkage
  values.
\item \textbf{Step 2.} The MA presents the certificate to the PCA and
  requests the hash of the individual certificate request from the
  PCA. The PCA delivers the hash of the individual certificate
  request, along with appropriate host information.
\item \textbf{Step 3.} Using the hash of the individual certificate
  request, the MA instructs the RA to add the enrollment certificate
  to its blacklist. Furthermore, the MA asks whether there are any
  non-expired certificates for this device.
\item \textbf{Step 4.} The RA replies with a list of hashes of
  individual certificate requests of non-expired certificates for the
  device to be revoked.
\item \textbf{Step 5.} Using the additional hashes of the individual
  certificate requests, the MA can retrieve the other non-expired
  certificates from the PCA.
\item \textbf{Step 6.} The MA adds CertIDs (a truncated hash of the
  certificate of, say, 8 bytes) of all non-expired certificates to the
  CRL.
\end{itemize}

Note that there are two possible optimizations of the process
described above. First, the PCA could return only the CertIDs of the
predecessor certificates and successor certificates in Step 5. Second,
Step 5 can be avoided if the RA chooses to store the CertID of
certificates not using linkage values (if certificates are not
encrypted by PCA to the device and RA is able to read certificates
provided to devices).

\subsection{CRL Size and CRL Distribution}
\label{sec:crl_size_and_distribution}
%
%
The size of the CRL grows linearly with the number of
revoked entities, and some devices will have less persistent storage
than others. It is assumed that all OEMs will provide at least enough 
storage for 10,000 entries, which translates to a file size of approximately 
400 KB. Therefore, a good CRL design will tag entries with information that
allows devices to identify the 10,000 (or more, depending on local storage)
entries that are of highest priority to them: for example, 
entries could be tagged with a location, or with the severity of misbehavior
associated with that device, or with an indicator that the private keys have
been made public. The final CRL design is still under development; a
preliminary design is provided in IEEE Std 1609.2-2016, but it does not
provide clear mechanisms for prioritization. 

Note that currently there is no way to undo a revocation, and a revoked device
can be reinstated only by repeating the process of bootstrapping,
cf.\ Section~\ref{sec:use_case_bootstrap}.

\subsubsection{Revocation and Tracking}
The value of having multiple linkage authorities is that it prevents an insider from gaining information that would enable him to track a vehicle, while at the same time allowing identification of a specific device under controlled circumstances. There are two circumstances where this is useful:

\begin{itemize}
\item  Revocation: as described above, the LAs enable efficient revocation via CRLs.
\item Misbehavior detection: If part of the misbehavior detection process is to check whether two messages, signed with different certificates, origin from the same device, then the LAs can be used to support this internal investigation in a privacy preserving manner. It is still a subject of research to determine what information LAs should be allowed to provide to MA, and under which circumstances.
\end{itemize}

\subsection{CRL Series: Identifying the Authorized Revoker for a Certificate}
\label{sec:crl_series}

Within the system there may be multiple \textit{CRL Sequences}, 
where a sequence is a temporally ordered set of CRLs addressing 
the same group of devices. The design ``pins'' each entity in the 
system to a specific CRL Sequence and allows different CRL sequences 
to be generated by different SCMS components.

There are a number of advantages to this:
\begin{itemize}
\item In the real world, jurisdiction over different vehicle or component types may rest in different authorities; it is therefore sensible to architect the system in such a way that devices do not have to be managed by a single central authority. This allows the appropriate authority to take responsibility for managing particular applications or fleets, avoiding a situation where a central CRL issuer might be reluctant to take responsibility for issuing CRLs for particular devices due to, for example, concerns about liability.
%
%
%
%
%
%
%
%
\item The natural revocation cycle length for different types of devices may be different. It may be appropriate to update the end-entity CRL daily, but the SCMS component CRL once a month. Alternatively, it might be desirable to publish CRLs continually as new revocation information becomes available -- this may be the case for CRLs for BSM senders as there may be a significant volume of revoked devices. Finally, some fleets, devices, or applications might hardly ever experience revocation. In this case they might not have regularly issued CRLs, but may instead issue a new CRL only when necessary.
%
%
%
%
%
%
%
%
\item Given that it makes sense to have different CRL Sequences, it is sensible to identify which CRL sequence is relevant to an entity:
\begin{itemize}
\item It improves efficiency of a revocation check when a message is received from that entity, as only the most recent CRL in only the relevant sequence need be checked;
\item It removes the risk that an identifier collision would result in an entity being accidentally revoked.
\item It ensures that if one CRL generator is compromised, it can only falsely revoke devices that it was supposed to have jurisdiction over, and cannot falsely revoke other devices. 
\end{itemize}
\end{itemize}

Note that although the ability to have different CRL signers is useful, if there are multiple CRL signers this potentially increases the communications burden on the vehicles as they now have to obtain multiple CRLs. This is addressed by having a central CRL store where all current CRLs are included in a single file which can be downloaded by vehicles as they obtain connectivity.

\begin{center}
\begin{figure*}[ht]
\scalebox{0.42}{
\input{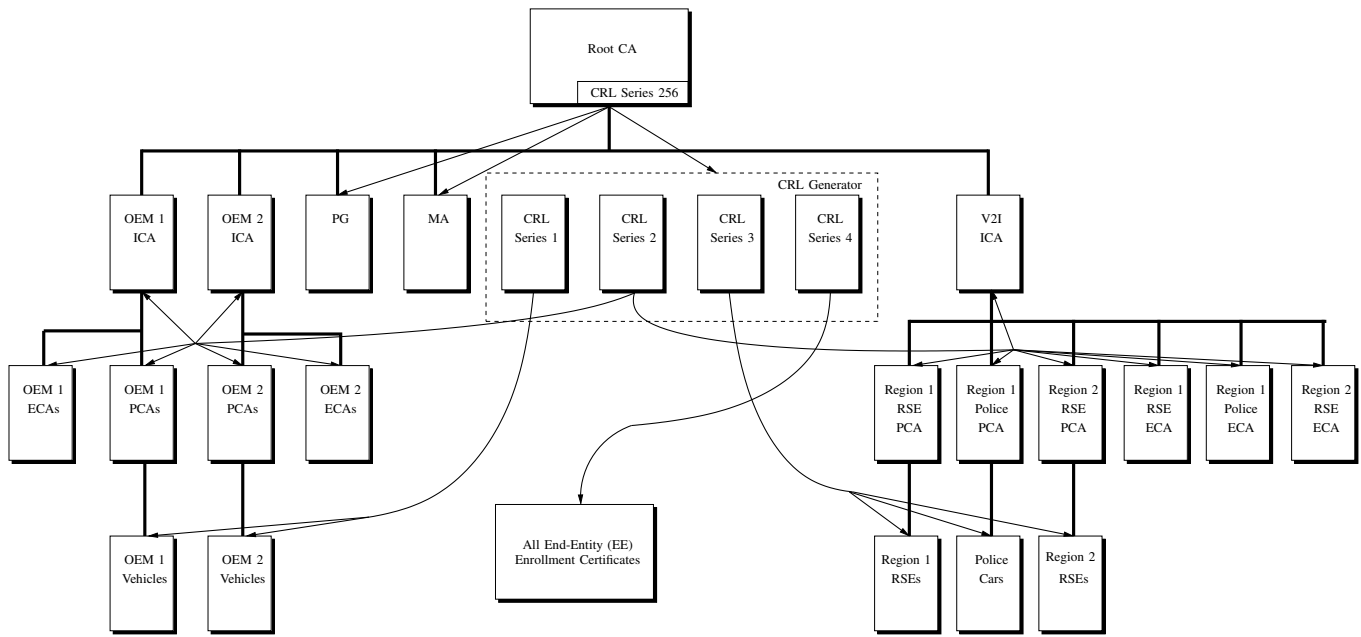}
}
\caption{\label{fig:crl_series}CRL Series}
\end{figure*}
\end{center}

This is done in the IEEE 1609.2 certificate and CRL structure by use of the \textit{Certificate Revocation Authority CA} (CRACA) ID and the CRL Series field. The CRACA is a CA in the chain of the potentially-revoked entity that either issues the CRL itself, or directly issues the CRL Generator certificate. The CRL Series appears in both the entity's certificate and the CRL and allows to distinguish between different CRL Sequences from the same CRL Generator. This provides unambiguous identification of the relevant CRL Sequence, similarly to (but more compactly than) the functionality provided by the CRLDistributionPoints in X.509 certificates.
%
%
%
%
%
%
Figure~\ref{fig:crl_series} shows the CRL series structure for the SCMS Proof of Concept. There is one main CRL generator. It manages four CRL Series, covering vehicle pseudonym certificates (series 1); all SCMS components except those identified below (series 2); vehicle identification and RSE application (series 3); and enrollment certificates (series 4). Additionally, the Root CA manages a CRL series, series 256, which can be used to revoke the Policy Generator, CRL Generator, and MA certificates. Note that for all of these CRL sequences, the Root CA acts as the CRACA, as it is the CA on the chain of the to-be-revoked devices; the CRL generator acts simply as an agent of the root CA and is kept separate so that the root CA can be kept offline to the greatest extent possible.
%
%
%
%

\section{CRL Processing and Distribution}
\label{sec:crl_processing}

To revoke all pseudonym certificates of a given device from a time period $i$ forward, both seeds $\ls_1(i)$ and $\ls_2(i)$, the linkage IDs $\laid_1$, $\laid_2$, and timing information $i$ and $j_{\mathrm{max}}$ are published on the CRL and distributed. A device that received the CRL forward hashes both seed values individually, calculates the pre-linkage values of the current time-period and then XORs the pre-linkage values to obtain the current linkage value. 
Figure~\ref{fig:linkage_value_revocation} shows the information which needs to be published for the revocation of a given device at time instant $i$, as well as the process for finding all possible linkage values for this device and $i' > i$.
Note that this
mechanism protects backward-privacy since certificates of revoked
devices cannot be identified for time-periods before the linkage seeds
were published.
%
%
%
%

\begin{center}
\begin{figure*}[ht]
\scalebox{1.1}{
\input{graphics/linkage_value_revocation.tex}
}
\caption{\label{fig:linkage_value_revocation}Revocation of Linkage Values}
\end{figure*}
\end{center}

For distribution of the CRL, we envision the use of ``collaborative'' distribution model
as defined in a  preliminary way by~\cite{hhl11}. In collaborative distribution, some devices
are initially seeded with CRLs (by RSEs, by cellular data, or by some other means)
and then distribute the CRLs to their peer devices as they drive past them in the
normal course of events. Devices that have received the CRL become distributors 
in their turn, allowing efficient coverage of the entire system with a small number
of initial seeders. Simulation results are promising, e.g.,~\cite{hhl11} shows that 
the area of Zurich can be provided with a CRL using a single RSE within a few hours.
However, an interesting open research question is how best to use relatively limited
(6Mbit/s) channel capacity in the 5.9 GHz band.
%

\section{Re-enrollment} \label{re-enrollment}
Re-enrollment of a device might be necessary due to several reasons. We define re-enrollment as either of the following: 
\begin{itemize}
\item \textbf{Reinstatement:} A device is reinstated if the original enrollment certificate is reinstated by removing it from the RA's blacklist.
\item \textbf{Re-bootstrapping:} A device is re-bootstrapped if the device is wiped and then bootstrap is executed to issue a new enrollment certificate. This is like a factory reset. 
\item \textbf{Re-issuance:} A device is reissued if the public key of the enrollment certificate is reused to issue a new enrollment certificate. The device keeps all pseudonym certificates and keeps using the same butterfly key parameters. 
\item \textbf{Re-establishment:} A device is re-established if the device's integrity can be verified remotely and the device then requests a new enrollment certificate using the old enrollment certificate to authenticate the request. 
\end{itemize}

Note that we strongly suggest to only use re-bootstrapping and re-establishment but no reinstating or re-issuing. Device re-enrollment is useful in the following scenarios:

\begin{itemize}
\item \textbf{Change of cryptography:} Advances in cryptanalysis might make it necessary to replace the underlying
cryptographic algorithms. In the next decades, this will likely be the case to introduce post-quantum
cryptography algorithms. In this case, devices need to receive updated firmware, ideally over-the-air, and
then request new enrollment certificates that use the updated cryptographic scheme. 
\item \textbf{Device revocation via CRL:} If a device is revoked and is listed on the CRL, only re-bootstrapping
is allowed.
\item \textbf{Enrollment certificate roll-over:} It is good practice and a security requirement in the SCMS to limit the
life-span of enrollment certificates, which motivates the need for a roll-over to a new enrollment 
certificate over-the-air, which is equivalent to re-establishing an EE. If the current enrollment certificate
has not been revoked, a device can request a new enrollment certificate. The device creates a new
 private/public key pair and includes that public key in its certificate roll-over request to RA. The device
digitally signs the roll-over request with its current enrollment certificate. The RA verifies the request,
forwards it to ECA and ECA in turn signs the requested enrollment containing the new public key. 
\item \textbf{Device revocation due to a revoked ECA:} If an ECA has been revoked such that
a device now holds an invalid enrollment certificate, re-enrollment is necessary as well. As
a standard approach, a device should be re-bootstrapped. Considering a re-establishment of devices that hold an enrollment certificate from a revoked ECA creates the risk to issue a new enrollment certificate to a malicious device, if there is no deterministic way to differentiate between malicious and honest devices. Hence we strongly recommend re-bootstrapping in a secure environment of all affected devices.
 
%
%
%
%
%
%
%
\item \textbf{Root CA and ICA revocation:} If a Root CA certificate is revoked, it is assumed that a new Root CA certificate is established by means of electors (see Section~\ref{sec:electors}) and all relevant components have been equipped with a new certificate under the new Root CA certificate. ECAs need to be re-certified, and
permission is given by the SCMS Manager to re-establish devices that hold an enrollment certificate issued
by a re-certified ECA, if there is evidence that the ECA has not been compromised. Otherwise devices need to be re-bootstrapped.

In case of either a Root CA certificate or ICA certificate revocation the revocation information will be distributed via an updated CRL. EEs will pick up the updated CRL, and check, if the certificate
is in the trust chain to their enrollment, pseudonym, application, or identification certificate. 
In case of the enrollment certificate an EE will send a re-enrollment request to the RA, using their current enrollment certificate. RAs receive information from the SCMS Manager, which ECAs are re-certified and will allow devices with an enrollment certificate signed by re-certified ECAs to request a new enrollment certificate. The EE needs to request new pseudonym, application, or identification certificates with the new enrollment certificate afterwards.

In case only the EE's pseudonym, application, or identification are affected by the revoked Root CA certificate, the EE needs to send new provisioning requests with the current enrollment certificate to the RA. RA will stop all pending pre-generation jobs with the previous PCA certificate and delete all pre-generated certificates as soon as it receives the updated CRL.

The case for a revoked ICA is equivalent, except that it's not necessary to create a new Root CA certificate and introduce it to the system via Electors.        
\end{itemize}

\section{Elector-based Root Management}
\label{sec:electors}
Given that EEs are able to re-enroll as described in subsection~\ref{re-enrollment} after a root CA certificate's validity period ends or a revocation was necessary and a new root CA certificate has been established for replacement, how can an EE start trusting this new root CA certificate? The trust in an initial root CA certificate is implicit, as it is installed in a secure environment with out-of-band communication during bootstrapping of the EE. One option would be to get the EE back to that secure environment and use out-of-band communication to install the new root CA certificate. But this is suboptimal due to the required effort and will render the overall V2X system partly out-of-order until all EEs have installed the new certificate.

To manage the root CA certificate over time and gain resilience against compromises on any level, the SCMS needs the ability to heal itself, which means to bring itself into a state where it can endure another singleton compromise or end of validity period. This recovery should occur while keeping the EEs operational whenever possible, that is, capable of sending, receiving and validating BSMs, and be able to heal the system hierarchy without requiring physical access to EEs. Elector-based Root Management as introduced in \cite{electors} is the solution that provides those means by installing a distributed management schema on top of the SCMS root CAs.

\subsection{Distributed Management, Electors}
A distributed management scheme, like a democracy, contains within itself the power to replace an established hierarchy, and does not succumb to a single failure: The concept of \textit{Electors}, which together have the power to change and manage the trust relationships of the system. The number of electors should be \(2n+1\), where n is the number of simultaneous elector expiration/compromises that can be tolerated.

Like in a democracy the Elector-based Root Management introduces a \textit{Ballot} with an \textit{Action} to endorse or revoke a root CA certificate or another elector. The electors cast \textit{Votes} by signing the ballot individually. When a quorum of valid electors voted on the ballot, the ballot can be trusted by any other component in the system and the action contained in the ballot can be executed securely.

The electors are not part of the PKI hierarchy, and therefore the electors can use a different crypto-system than the SCMS PKI, in fact each of them can use a different one. This raises the probability that in case of a root CA certificate compromise due to a broken cryptography, the system is still able to heal itself. 

The resulting system may have multiple, self-signed root CA certificates, each of which operates as the top of their trust chain. Each root CA's certificate is endorsed by a ballot with at least a quorum of votes from non-revoked electors. EEs need to verify the trust chain up to a root CA certificate, at which point they must verify that a quorum of non-revoked electors have endorsed that root CA certificate.

\subsection{Ballots \& Actions}
Electors operate by signing ballot messages to be consumed by other SCMS components, including EEs. These ballots include the following basic set of actions:

Add:
\begin{itemize}
\item Endorse root CA certificate
\item Endorse elector certificate
\end{itemize}

Revoke:
\begin{itemize}
\item Revoke root CA certificate
\item Revoke elector certificate
\end{itemize}

Each ballot contains only one action, the certificate that action should be applied to, and multiple signatures of electors. Components, including EEs, know the quorum and the certificates of the initial set of electors and therefore can validate the authorization of the action contained in the ballot.

The SCMS Manager will coordinate the production of these ballot messages.

\subsection{Structure of Ballots}
Each elector shall produce an independent vote in support of a root or elector endorsement or in support of a root or elector revocation. All independent Elector votes will be collected in one ASN.1 structure which is referred to as a Ballot. The ballot structure shall contain the following elements:
\begin{enumerate}
\item A Sequence of Actions, each containing:
\begin{enumerate}[label=\alph*)]
\item The Certificate of the \textbf{Object} of the Action
\item A sequence of Elector \textbf{Signatures}.
\end{enumerate}
\end{enumerate}
Note that the validity period of a ballot is implicitly given by the validity period of the object of the action. 

\subsection{Revocation/Endorsement Impact on EEs}
A key consideration in the design of the root management system is to maintain secure operation of EEs without requiring recall or manual re-enrollment of individual devices (as described in section ~\ref{re-enrollment}). Table~\ref{table:ee_impact} outlines the status of EEs through the addition or revocation of electors and root CAs.

\begin{table*}[ht]
\caption{Root Management Message Impact on EEs}
\centering
\begin{tabular}{|l|p{0.75\textwidth}|}
\hline
Action & Impact on EE operations\\ 
\hline \hline
\multirow{7}{*}{Revocation of an Elector} & 
As long as there are at least three electors with a quorum of two, then one elector may be removed without impacting operation: The remaining electors are still a quorum and their endorsements of the root CA certificate would still be valid. A single revoked elector would not stop operations of any EE. A replacement elector may then be added back to the system to return to a state with three valid electors. A larger number of electors may be used to improve the system's resilience to compromise or failure of these top-level trust anchors. \\ \hline

\multirow{5}{*}{Revocation of a Root CA} & 
Revoking a root CA certificate would stop operations of EEs that possess certificates chaining up to the revoked root CA certificate. Those EEs would need to re-enroll and be re-provisioned with a different root CA before they could be trusted by other devices.\\ \hline

\multirow{6}{*}{Addition of an Elector} & 
A new self-signed elector certificate that is endorsed by a quorum of valid electors can be trusted by EEs and other SCMS components without the need of returning them to a secure environment.

In addition, this new elector can endorse existing root CA certificates without the need for any updates of the existing valid certificates, including the EE's pseudonym certificates.\\ \hline

\multirow{3}{*}{Addition of a Root CA} & 
A new, self-signed root CA certificate that is endorsed by a quorum of valid electors can be trusted by EEs and other SCMS components without the need of returning them to a secure environment. EEs can immediately begin to trust messages that chain up to the new root CA.\\ \hline
\end{tabular}
\label{table:ee_impact}
\end{table*}

%
%

\section{Organizational Separation}
\label{sec:organizational_separation}
Different SCMS components represent different logical functions. For
providing an acceptable level of privacy for V2V safety communication applications using BSMs
%
%
and pseudonym certificates, some distinct logical functions must be
provided by distinct organizations, in order to prevent a single organization
from being able to determine which pseudonym certificates belong to a
device. This capability would allow an attacker to track a vehicle by 
capturing its BSMs. 

This section identifies which SCMS
components must be organizationally separate.  The general rule is
that two components cannot be run by the same organization if the
combined information held by the components would allow an insider to
determine which pseudonym certificates belong to a
device. This leads to the following specific requirements for organizational separation:

\begin{itemize}
\item \textbf{PCA and RA:} If these two components were run by one
  organization, the organization would know which pseudonym
  certificates had been issued to which device. This is because the RA
  knows the requests to which certificates correspond, and the PCA
  knows the corresponding pseudonym certificates.

\item \textbf{PCA and one of the LAs:} If an organization ran the PCA
  and either (or, both) of the LAs, it could link all pseudonym
  certificates (from any batch) issued to any device since LA
  knows a set of pre-linkage values that go into the certificate set,
  and PCA sees these pre-linkage values at certificate generation
  time.

\item \textbf{LA$_1$ and LA$_2$:} If an organization ran both the LAs,
  it would know all the pre-linkage values and XOR them
  opportunistically to obtain the linkage values, which appear in
  plaintext in pseudonym certificates. This would allow identification of which   
  pseudonym certificates belong to the same device.

\item \textbf{LOP and (RA or MA):} The LOP hides the
  location from the RA and the MA, respectively, and may not be
  combined with any of these components.

\item \textbf{MA and (RA, LA, or PCA):} The MA should not be combined
  with any of the RA, the LA or the PCA. If combined, the MA could
  circumvent restrictions during misbehavior investigation and learn 
  more information than necessary for misbehavior investigation and 
  revocation purposes.
\end{itemize}

When other certificate types are generated, no specific organizational separation is required.

%
%

\section{Conclusions and Outlook}
\label{sec:conclusions}

We introduced a Security Credential Management System (SCMS) for V2X
communications, with a special emphasis on V2V safety application
communication. This SCMS design is a leading candidate for the V2X
security backend design in the US. One of the remaining challenges is to define policies that 
balance among security, privacy, and efficiency that will support the establishment of a 
nationwide system. The proposed solution uses five certificate types to
%
%
cover all identified V2X application categories. This article focused
on V2V safety communication security and the life-cycle of OBE
pseudonym certificates since this certificate type requires most
consideration in terms of privacy protection and complexity. The SCMS
is designed to scale with the number of devices and to protect privacy
of end-users against inside and outside attackers by separation of
duties.

Next steps towards an SCMS deployment are as follows:

\begin{itemize}
\item Define an effective and CRL dissemination based on the concept of Collaborative
  Distribution~\cite{hhl11}.
\item Large scale test deployments of a proof-of-concept SCMS.
\item Development of misbehavior detection algorithms, implementation and
test of misbehavior reporting, investigation and detection in a proof-of-concept
implementation.
\end{itemize}

\section*{Acknowledgments}

The presented results are a culmination of efforts by many parties and
people. This includes members of the US Department of Transportation
(USDOT), the Crash Avoidance Metrics Partners LLC Vehicle Safety
Consortium (CAMP VSC3, CAMP VSC5), and the Vehicle Infrastructure Integration
Consortium (VIIC). Funding was provided by the USDOT and OEMs
participating in CAMP VSC3 and VSC5. The OEMs Daimler, Ford, GM, Honda,
Hyundai-Kia, Nissan, Toyota, and VW-Audi participate in CAMP VSC3 and
the VIIC. BMW and Chrysler participate in the VIIC.  The authors are
deeply grateful to Bill Anderson, Stephen Farrell, the late Scott Vanstone, and
members of the USDOT and CAMP VSC3 for reviewing earlier versions of
the manuscript and providing important suggestions for improvements.

\bibliographystyle{IEEEtran}
\bibliography{ms}

\begin{thebibliography}{10}
\providecommand{\url}[1]{#1}
\csname url@samestyle\endcsname
\providecommand{\newblock}{\relax}
\providecommand{\bibinfo}[2]{#2}
\providecommand{\BIBentrySTDinterwordspacing}{\spaceskip=0pt\relax}
\providecommand{\BIBentryALTinterwordstretchfactor}{4}
\providecommand{\BIBentryALTinterwordspacing}{\spaceskip=\fontdimen2\font plus
\BIBentryALTinterwordstretchfactor\fontdimen3\font minus
  \fontdimen4\font\relax}
\providecommand{\BIBforeignlanguage}[2]{{%
\expandafter\ifx\csname l@#1\endcsname\relax
\typeout{** WARNING: IEEEtran.bst: No hyphenation pattern has been}%
\typeout{** loaded for the language `#1'. Using the pattern for}%
\typeout{** the default language instead.}%
\else
\language=\csname l@#1\endcsname
\fi
#2}}
\providecommand{\BIBdecl}{\relax}
\BIBdecl

\bibitem{usdot-cv}
\BIBentryALTinterwordspacing
{U.S. Department of Transportation}. {How Connected Vehicles Work}. [Online].
  Available: \url{http://www.its.dot.gov/factsheets/pdf/JPO_HowCVWork_v3.pdf}
\BIBentrySTDinterwordspacing

\bibitem{usdot-2017}
\BIBentryALTinterwordspacing
{U.S. Department of Transportation - National Highway Traffic Safety
  Administration}. {U.S. DOT Federal Motor Vehicle Safety Standards; V2V
  Communications; NPRM}. [Online]. Available:
  \url{https://www.federalregister.gov/documents/2017/01/12/2016-31059/federal-motor-vehicle-safety-standards-v2v-communications}
\BIBentrySTDinterwordspacing

\bibitem{c2c11}
N.~Bi{\ss}meyer, H.~St\"ubing, E.~Schoch, S.~G\"otz, J.~P. Stolz, and B.~Lonc,
  ``{A generic public key infrastructure for securing Car-to-X
  communication},'' in \emph{18th World Congress on Intelligent Transport
  Systems}, 2011.

\bibitem{etsi-tr}
\BIBentryALTinterwordspacing
{ETSI}. {TR 102 893V1.1.1 (2010-03) Intelligent Transport Systems (ITS);
  Security; Threat, Vulnerability and Risk Analysis (TVRA)}. [Online].
  Available:
  \url{http://www.etsi.org/deliver/etsi_tr/102800_102899/102893/01.01.01_60/tr_102893v010101p.pdf}
\BIBentrySTDinterwordspacing

\bibitem{etsi-ts1}
\BIBentryALTinterwordspacing
------. {TS 102 731V1.1.1 (2010-09) Intelligent Transport Systems (ITS);
  Security; Security Services and Architecture}. [Online]. Available:
  \url{http://www.etsi.org/deliver/etsi_ts/102700_102799/102731/01.01.01_60/ts_102731v010101p.pdf}
\BIBentrySTDinterwordspacing

\bibitem{etsi-ts2}
------. {TS 102 867 v1.1.1, Intelligent Transportation Systems (ITS); Security;
  Stage 3 mapping for IEEE 1609.2}.

\bibitem{sevecom}
\BIBentryALTinterwordspacing
{Secure Vehicle Communication}. {Security Architecture and Mechanisms for
  V2V/V2I}. [Online]. Available:
  \url{https://sevecom.eu/Deliverables/Sevecom_Deliverable_D2.1_v3.0.pdf}
\BIBentrySTDinterwordspacing

\bibitem{vscc}
{Vehicle Safety Communications Consortium}. {Appendix H: Final Report 2006}.

\bibitem{1609.2}
\BIBentryALTinterwordspacing
{IEEE 1609.2}. {Annex E.4.1: Why sign data instead of using a message
  authentication code?} [Online]. Available:
  \url{https://standards.ieee.org/findstds/standard/1609.2-2013.html}
\BIBentrySTDinterwordspacing

\bibitem{emvco}
\BIBentryALTinterwordspacing
{EMVCo, LLC}. {Worldwide EMV Deployment Statistics}. [Online]. Available:
  \url{https://www.emvco.com/about_emvco.aspx?id=202}
\BIBentrySTDinterwordspacing

\bibitem{wiki:pki}
\BIBentryALTinterwordspacing
Wikipedia, ``Public key infrastructure --- wikipedia{,} the free
  encyclopedia,'' 2016, [accessed 28-February-2017]. [Online]. Available:
  \url{https://en.wikipedia.org/w/index.php?title=Public_key_infrastructure&oldid=755924137}
\BIBentrySTDinterwordspacing

\bibitem{safety-pilot}
\BIBentryALTinterwordspacing
{U.S. Department of Transportation}. {Safety Pilot Model Deployment}. [Online].
  Available: \url{http://safetypilot.umtri.umich.edu}
\BIBentrySTDinterwordspacing

\bibitem{scms-interface-spec}
\BIBentryALTinterwordspacing
{Crash Avoidance Metrics Partners LLC}. Scms interface specification. [Online].
  Available:
  \url{https://stash.campllc.org/projects/SCMS/repos/scms-asn/browse}
\BIBentrySTDinterwordspacing

\bibitem{wwkh13}
\BIBentryALTinterwordspacing
W.~Whyte, A.~Weimerskirch, V.~Kumar, and T.~Hehn, ``A security credential
  management system for v2v communications,'' \emph{2013 {IEEE} Vehicular
  Networking Conference (VNC 2013)}, 2013. [Online]. Available:
  \url{http://ieeexplore.ieee.org/document/6737583}
\BIBentrySTDinterwordspacing

\bibitem{j2945.1}
\BIBentryALTinterwordspacing
{SAE}. (2016) {J2945/1\_201603, On-Board System Requirements for V2V Safety
  Communications}. [Online]. Available:
  \url{http://standards.sae.org/j2945/1_201603/}
\BIBentrySTDinterwordspacing

\bibitem{PfitzmannK00}
A.~Pfitzmann and M.~K{\"{o}}hntopp, ``Anonymity, unobservability, and
  pseudonymity - {A} proposal for terminology,'' in \emph{Designing Privacy
  Enhancing Technologies, International Workshop on Design Issues in Anonymity
  and Unobservability, Berkeley, CA, USA, July 25-26, 2000, Proceedings}, 2000,
  pp. 1--9.

\bibitem{RFFingerprinting}
V.~Brik, S.~Banerjee, M.~Gruteser, and S.~Oh, ``Wireless device identification
  with radiometric signatures,'' in \emph{MOBICOM}, 2008, pp. 116--127.

\bibitem{hhl11}
J.~J. Haas, Y.-C. Hu, and K.~P. Laberteaux, ``Efficient certificate revocation
  list distribution,'' \emph{{IEEE} Journal of Selected Areas in Communications
  - Special Issue on Vehicular Communications and Networks}, vol. 29(3), pp.
  595--604, January 2011.

\bibitem{tls}
\BIBentryALTinterwordspacing
T.~Dierks and E.~Rescorla. (2008, August) Rfc 5246 - the transport layer
  security tls protocol version 1.2. [Online]. Available:
  \url{http://tools.ietf.org/html/rfc5246}
\BIBentrySTDinterwordspacing

\bibitem{sybil}
J.~R. Douceur, ``The sybil attack,'' in \emph{IPTPS}, 2002, pp. 251--260.

\bibitem{PetitSFK15}
J.~Petit, F.~Schaub, M.~Feiri, and F.~Kargl, ``Pseudonym schemes in vehicular
  networks: {A} survey,'' \emph{{IEEE} Communications Surveys and Tutorials},
  vol.~17, no.~1, pp. 228--255, 2015.

\bibitem{kw11}
\BIBentryALTinterwordspacing
H.~Krishnan and A.~Weimerskirch, ``Verify-on-demand - a practical and scalable
  approach for broadcast authentication in vehicle-to-vehicle communication,''
  \emph{SAE 2011 World Congress}, 2011. [Online]. Available:
  \url{http://papers.sae.org/2011-01-0584}
\BIBentrySTDinterwordspacing

\bibitem{ieee1363}
\BIBentryALTinterwordspacing
{IEEE}. (2000) Std 1363-2000, standard specifications for public-key
  cryptography. [Online]. Available:
  \url{https://ieeexplore.ieee.org/xpl/articleDetails.jsp?arnumber=891000&contentType=Standards}
\BIBentrySTDinterwordspacing

\bibitem{NistCurves}
\BIBentryALTinterwordspacing
{National Institute of Standards and Technology}. (1999, July) {Recommended
  elliptic curves for federal government use}. [Online]. Available:
  \url{http://csrc.nist.gov/groups/ST/toolkit/documents/dss/NISTReCur.doc}
\BIBentrySTDinterwordspacing

\bibitem{implicit11}
SEC, \emph{{Elliptic Curve Qu-Vanstone Implicit Certificate Scheme (ECQV)}},
  SEC Std. 4, 2011.

\bibitem{scms-ee-ma-interface}
\BIBentryALTinterwordspacing
{Crash Avoidance Metrics Partners LLC}. Ee-ma interface specification.
  [Online]. Available:
  \url{https://stash.campllc.org/projects/SCMS/repos/scms-asn/browse/ee-ma.asn}
\BIBentrySTDinterwordspacing

\bibitem{electors}
\BIBentryALTinterwordspacing
B.~Brecht, D.~Therriault, R.~Motz, V.~Kumar, R.~Lambert, W.~Lattin,
  B.~Romansky, and W.~Whyte. (2016, March) Elector-based root management system
  to manage a public key infrastructure. [Online]. Available:
  \url{http://priorart.ip.com/IPCOM/000245336}
\BIBentrySTDinterwordspacing

\end{thebibliography}
\end{document}